\documentclass[aps,reprint,groupedaddress,showpacs]{revtex4-1}
\usepackage{amsmath,bm}
\usepackage{amssymb}
\usepackage{color}
\usepackage{multirow}
\usepackage[colorlinks,bookmarks=false,citecolor=red,linkcolor=blue,urlcolor=blue]{hyperref}
\usepackage{graphicx}
\usepackage{mathrsfs,bbm}
\usepackage[caption=false]{subfig} 
\usepackage{wrapfig}
\usepackage[version=3]{mhchem}

\def\Sbf{{\bf S}}
\def\m{{\bf m}}

\def\a{{\bf a}}
\def\b{{\bf b}}
\def\k{{\bf k}}
\def\r{{\bf r}}
\def\q{{\bf q}}
\def\G{{\bf G}}

\def\onebar{\bar{1}}
\def\sbar{\bar{s}}
\def\Hhat{\hat{H}}
\def\Phat{\hat{P}}
\def\Qhat{\hat{Q}}
\def\bhat{\hat{b}}
\def\shat{\hat{s}}
\def\that{\hat{t}}
\def\xhat{\hat{x}}
\def\yhat{\hat{y}}

\begin{document}
\title{Plaquette-triplon analysis of magnetic disorder and order in a trimerized spin-1 Kagom\'e antiferromagnet}
\author{Pratyay Ghosh}
\author{Akhilesh Kumar Verma}
\author{Brijesh Kumar}
\email{bkumar@mail.jnu.ac.in}
\affiliation{School of Physical Sciences, Jawaharlal Nehru University, New Delhi 110067, India}
\date{\today}


\begin{abstract}
A spin-1 Heisenberg model on trimerized Kagom\'e lattice is studied by doing a low-energy bosonic theory in terms of plaquette-triplons defined on its triangular unit-cells. The model considered has an intra-triangle antiferromagnetic exchange interaction, $J$ (set to 1), and two inter-triangle couplings, $J^\prime>0$ (nearest-neighbor) and $J^{\prime\prime}$ (next-nearest-neighbor; of both signs). The triplon analysis performed on this model investigates the stability of the trimerized singlet ground state (which is exact in the absence of inter-triangle couplings) in the $J^\prime$-$J^{\prime\prime}$ plane. 
It gives a quantum phase diagram that has two gapless antiferromagnetically ordered phases separated by the spin-gapped trimerized singlet phase. The trimerized singlet ground state is found to be stable on $J^{\prime\prime}=0$ line (the nearest-neighbor case), and on both sides of it for $J^{\prime\prime}\neq 0$, in an extended region bounded by the critical lines of transition to the gapless antiferromagnetic phases. The gapless phase in the negative $J^{\prime\prime}$ region has a coplanar $120^\circ$-antiferromagnetic order with $\sqrt{3}\times\sqrt{3}$ structure. In this phase, all the magnetic moments are of equal length, and the angle between any two of them on a triangle is exactly $120^\circ$. The magnetic lattice in this case has a unit-cell consisting of three triangles. The other gapless phase, in the positive $J^{\prime\prime}$ region, is found to exhibit a different coplanar antiferromagnetic order with ordering wavevector $\q=(0,0)$. Here, two magnetic moments in a triangle are of same magnitude, but shorter than the third. While the angle between  two short moments is $120^\circ-2\delta$, it is $120^\circ+\delta$ between a short and the long one. Only when $J^{\prime\prime}=J^\prime$, their magnitudes become equal and the relative-angles $120^\circ$. The magnetic lattice in this $\q=(0,0)$ phase has the translational symmetry of the Kagom\'e lattice with triangular unit-cells of reduced (isosceles) symmetry. This reduction in the point-group symmetry is found to show up as a difference in the intensities of certain Bragg peaks, whose ratio, $I_{(1,0)}/I_{(0,1)} = 4\sin^2{(\frac{\pi}{6}+\delta)}$, presents an experimental measure of the deviation, $\delta$, from the $120^\circ$ order.
\end{abstract}

\maketitle

\section{\label{sec:intro} Introduction}
The quantum antiferromagnets on frustrated lattices, with competing interactions, tend to disfavor magnetic ordering, and realize interesting quantum-disordered low temperature phases (ground states) such as the quantum spin liquids, valence-bond-solid states, or the dimer or plaquette ordered singlet phases~\cite{Caspers,Indrani,Misguich2012,Balents2010}. The Kagom\'e quantum antiferromagnet is an interesting example of a frustrated spin system, in which the frustrated geometry of the Kagom\'e lattice (a triangular lattice of the corner-sharing triangles) and the quantum fluctuations together present a serious detriment to magnetic ordering in its ground state. For instance, the low temperature properties of \ce{Cu3Zn(OH)6Cl2}~{\cite{Mendels2008,Mendels2011}}, \ce{BaCu3V2O8(OH)2}~{\cite{Okamoto2009}}, \ce{[NH4]2[C7H14N][V7O6F18]}~{\cite{Aidoudi2011}} and \ce{\gamma-Cu3Mg(OH)6Cl2}~{\cite{Colman2011}}, which are the realizations of the spin-$1/2$ Kagom\'e Heisenberg antiferromagnet (KHA), seem to indicate this. While there is a strong support for the spin-1/2 KHA with nearest-neighbor interactions to have a spin liquid ground state, the  difficult nature of this problem has made it very  hard to settle the debate on the true character of its ground state~{\cite{Zeng1990, Marston1991, Chalker1992, Leung1993, Subra1995, Lecheminant1997, Mila1998, Sindzingre, Zhitomirsky2004, Singh2008, Evenbly2010, Hwang2011, Li2012, Iqbal2013, Clark2013}}. The relatively lesser studied spin-1 and higher spin KHA's are also not very well understood.

For the spin-1 antiferromagnetic Heisenberg model on Kagom\'e lattice, which is the problem of interest to us in the present paper, a non-magnetic hexagonal-singlet-solid (HSS) ground state with gapped magnetic excitations was proposed by Hida using the exact diagonalization and cluster expansion methods~{\cite{Hida2000}}. A more recent study of the spin-1 KHA model using coupled-cluster method also found a non-magnetic ground state~{\cite{Gotze2011}}. This problem is currently in a surge of theoretical activity, motivated by the experiments on several spin-1 Kagom\'e materials~\cite{Wada1997, Matsushita2010, Hara2012, Behera2006, Papoutsakis2002}. While some of these materials do not neatly qualify as spin-1 Kagom\'e antiferromagnets (due to their ferromagnetic or glassy response), but there are some, e.g. \ce{m-MPYNN.BF4}~{\cite{Wada1997,Matsushita2010}} and \ce{KV3Ge2O9}~{\cite{Hara2012}}, which clearly show frustrated antiferromagnetic behavior. While \ce{m-MPYNN.BF4} is well-known to be spin-gapped at low temperatures, the behavior of \ce{KV3Ge2O9} is reported to be more exotic. Another material, \ce{NaV6O11} (a metallic vanadate), has three types of vanadium ions, of which one type forms the spin-1 Kagom\'e layers exhibiting spin-gap behavior below 243K, accompanied by explicit trimerization~\cite{Kato2001, Uchida2001}. The appearance of weak spontaneous magnetization below 65K (due to other vanadium ions), however, undermines the trimerized singlet physics of its Kagom\'e layers.

The most recent numerical  calculations using tensor network algorithms~\cite{Liu2015,Picot2015}, exact diagonalization and density matrix renormalization group (DMRG)~\cite{Changlani2015} find a gapped spontaneously trimerized singlet ground state for the quantum spin-1 nearest-neighbor KHA model. There are others who either support the HSS state of Hida~\cite{Nishimoto2015.HSS}, or suggest a gapped resonating AKLT (Affleck-Kennedy-Lieb-Tasaki) loop ground state~\cite{Li2014.RAL}. Despite the differences, they all point towards a spin-gapped non-magnetic ground state for the spin-1 KHA with nearest-neighbor interaction, which is in clear contrast with the studies that predicted $\sqrt{3}\times\sqrt{3}$ antiferromagnetic order in its ground state~\cite{Xu2007,Chubukov1992}. 
The spin-1 Kagom\'e antiferromagnet with spin anisotropies and biquadratic interaction have also been investigated~\cite{Damle2006,Xu2007}, but the pure Heisenberg case is what concerns us presently. Beyond the nearest-neighbor case, the trimerized singlet ground state has also been discussed for a spin-1 KHA with certain specific second and third neighbor interactions~\cite{Cai2009}.

Motivated by these recent studies on the ground state of spin-1 KHA, we investigate in this paper a Heisenberg model, described in Sec.~\ref{sec:model}, on trimerized Kagom\'e lattice. Our basic idea and the strategy are as follows. Since the Kagom\'e lattice is a triangular lattice of corner-sharing triangles, we construct an effective theory of spin-1 KHA in terms of the eigenstates of its basic triangular units. This we do by deriving, in Sec.~\ref{subsec:spinoperator}, a bosonic representation for the spin-1 operators of a triangular plaquette in terms of its singlet and triplet states. It is like the bond-operator representation of the spins of a dimer~\cite{Sachdev1990,Kumar2010}. This effective theory, formulated in Sec.~\ref{subsec:triplon-mft}, allows us to study the stability of the trimerized singlet (TS) ground state with respect to the elementary triplon (dispersing triplet) excitations, and to find if there is any antiferromagnetic (AF) order. Unlike the spin-wave analysis, which is a small fluctuation bosonic theory for a given classical magnetic order, this plaquette-triplon theory is formulated with respect to the non-magnetic TS state which is `quantum disordered'. It can describe classical order as well as quantum disorder in the ground state. 

From the triplon analysis performed in this paper, we find a stable TS ground state for the nearest-neighbor spin-1 KHA, in agreement with recent numerical studies~\cite{Liu2015, Picot2015, Changlani2015}. We also find this gapped TS phase over a range of second-neighbor interaction. Eventually, for sufficiently negative second-neighbor interaction, it undergoes a transition to the gapless phase with coplanar $120^\circ$-AF order with $\sqrt{3}\times\sqrt{3}$ structure, in which the neighboring magnetic moments lie at $120^\circ$ angle relative to each other, and the magnetic unit-cell consists of three triangles. This AF order has been known to occur in the KHA model for large spins with second and third neighbor interactions~\cite{Harris1992,Chubukov1992}. But here, we find it for spin-1, emerging spontaneously from the quantum disordered TS state. For positive second-neighbor interaction, we find a different coplanar AF order with ordering wavevector $\q=(0,0)$. In this phase, the magnetic moments in a triangular unit-cell are of unequal magnitudes (two short and one long), and at a deviation of $\delta$ from the $120^\circ$ orientation (with $120^\circ-2\delta$ angle between the short moments, and $120^\circ+\delta$ between the long and the short ones). This AF order has not been discussed before in Kagom\'e antiferromagnets, but here it emerges spontaneously. We discuss these findings in detail in Sec~\ref{sec:discussion}, and conclude this work with a summary in Sec~{\ref{sec:sum}}.
 
\section{\label{sec:model} Model}
\begin{figure}[t]
 \centering
 \includegraphics[width=0.9\columnwidth]{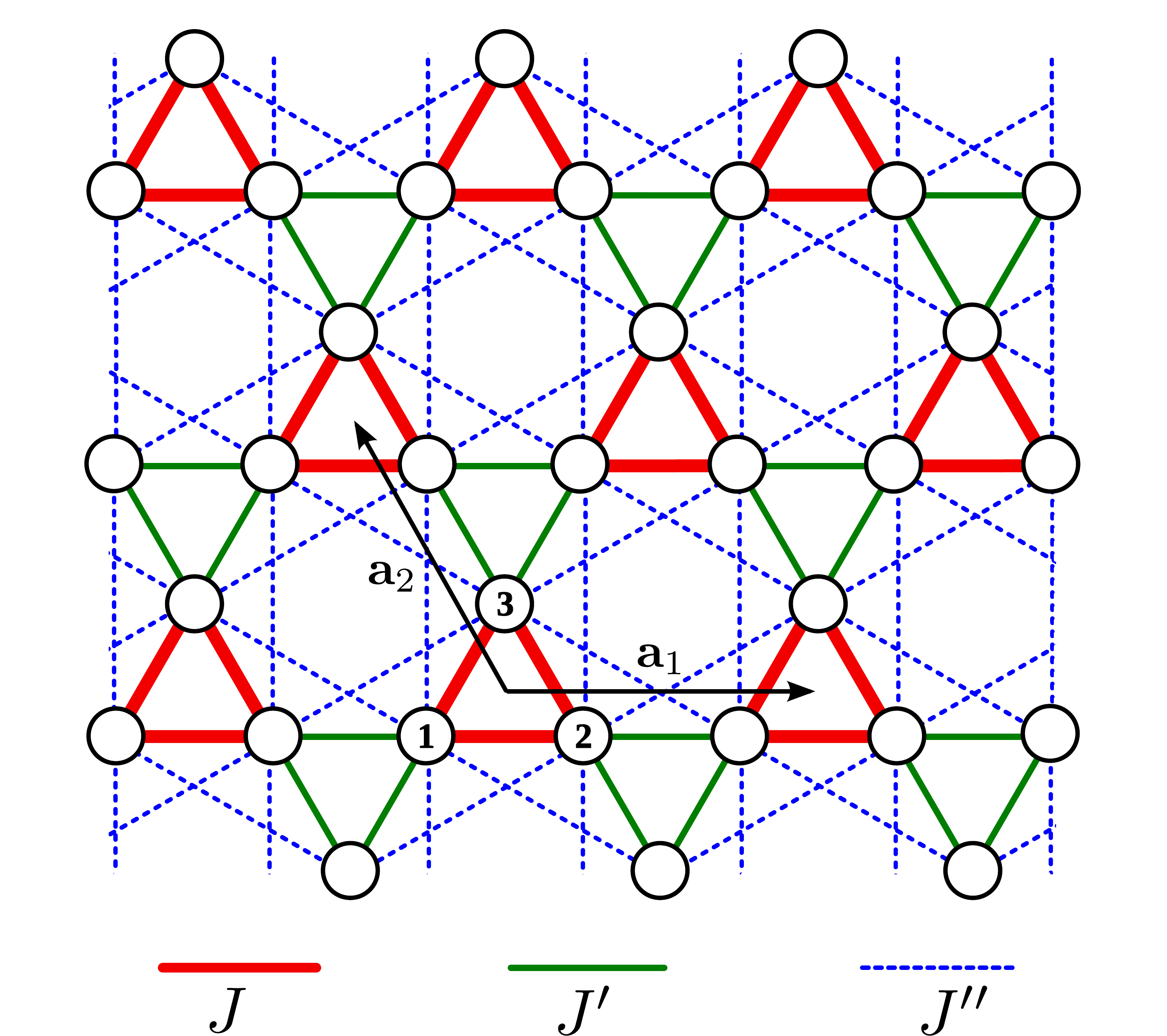}
 \caption{The trimerized Kagom\'{e} lattice with intra-triangle Heisenberg exchange interaction, $J$ (thick-red), and the inter-triangle exchange couplings, $J^\prime$ (green) and $J^{\prime\prime}$ (dashed-blue). The $\a_1=2\xhat$ and $\a_2=-\xhat + \sqrt{3}\yhat$ are two primitive vectors of this lattice.}
 \label{fig:TKagome}
\end{figure}

In this paper, we study the following quantum spin-1 Hamiltonian on trimerized Kagom\'e lattice. 
\begin{equation}{\label{eq:H}}
 \hat{H}=J\sum_{\langle i,j \rangle}\vec{S}_{i}\cdot\vec{S}_{j}+J^{\prime}\sum_{\left(i,j\right) }\vec{S}_{i}\cdot\vec{S}_{j}+J^{\prime\prime}\sum_{\langle\langle i,j \rangle\rangle}\vec{S}_{i}\cdot\vec{S}_{j}
\end{equation}
As depicted in Fig.~\ref{fig:TKagome}, it is a problem of the coupled antiferromagnetic triangles. In this explicitly trimerized Kagom\'e problem, the exchange interaction, $J$, in the up (red) triangles is taken to be stronger than that in the down (green) triangles, $J^\prime$. It resembles, \emph{for instance}, the trimerized Kagom\'e layers of one type of vanadium ions in \ce{NaV6O11}~\cite{Kato2001}. This similarity is only partial, however, as the situation in \ce{NaV6O11} is a bit more complex (and not to be dwelt upon here). We also include second-neighbor interaction, $J^{\prime\prime}$, for generality. In Eq.~(\ref{eq:H}), ${\langle i,j \rangle}$ denotes the spin-pairs in the up triangles, $\left( i,j \right)$ denotes the spin-pairs in the down triangles, and the second-neighbor spin-pairs are denoted as $\langle\langle i, j \rangle\rangle$. We take $J$ and $J^\prime$ to be antiferromagnetic, and allow $J^{\prime\prime}$ to be positive as well as negative. The $\hat{H}$ becomes the standard nearest-neighbor KHA for $(J^\prime,J^{\prime\prime})=(J,0)$.

A simple limiting case of the $\hat{H}$ of Eq.~(\ref{eq:H}) corresponds to $(J^\prime,J^{\prime\prime})=(0,0)$, for which the exact ground state is given by the direct product, $\otimes\prod^{ }_{\bigtriangleup} |s\rangle$, of the singlet states, $|s\rangle$, of the up-triangles. It also has an energy-gap, $J$, to the triplet excitations. Refer to Appendix~\ref{sec:triangle} for the eigenstates of spin-1 Heisenberg model on a triangle. Since three spin-1's uniquely form a singlet, this \emph{ideal} spin-gapped trimerized singlet ground state is also unique. 
How the non-zero $J^\prime$ and $J^{\prime\prime}$ affect this non-magnetic TS ground state is the question that we try to answer here by studying its stability against the triplon excitations.

\section{\label{sec:MFT} Triplon analysis}
Just as in the case of dimerized quantum antiferromagnets, where the bond-operator formalism provides a convenient means to construct an effective low-energy theory~\cite{Sachdev1990,Kumar2008}, here we do a low-energy theory of the trimerized KHA model, $\hat{H}$, in terms of the plaquette-operators defined on the triangular unit-cells of Kagom\'e lattice. Towards this goal, we first derive a bosonic representation for the spin-1 operators of a triangle, and then formulate a simple but useful theory of $\hat{H}$ in terms of these plaquette-operators. This theory will find for us the region of stability of the TS state, and identify the magnetic order, if any, in the $J^\prime$-$J^{\prime\prime}$ plane.

\subsection{\label{subsec:spinoperator} Plaquette-Operator Representation of the Spin-1 Operators on a Triangle}
The spin-1 Heisenberg model on a triangle, that is $\hat{H}_\triangle = J(\Sbf_1\cdot\Sbf_2 + \Sbf_2\cdot\Sbf_3 +\Sbf_1\cdot\Sbf_3)$, has a unique singlet eigenstate, $|s\rangle$, with eigenvalue, $-3J$. It has three sets of triplets (that is, 9 degenerate states with total-spin equal to 1), $|t_{m\nu}\rangle$, given by $m=1,0,\onebar$ (total-$S_z$) and $\nu=1,0,\onebar$. Here, $-1$ is denoted as $\onebar$, and the quantum number $\nu$ comes from the threefold rotational symmetry of $\Hhat_\triangle$. The energy of these triplets is $-2J$. It also has two sets of quintets (10 states with total-spin 2) and a heptet (7 eigenstates with total-spin 3) with energies 0 and $3J$, respectively. The eigenvalue problem for $\Hhat_\triangle$ is worked out in detail in Appendix~\ref{sec:triangle}. 

For $J>0$, the singlet is the ground state of $\Hhat_\triangle$ and the triplets form the elementary excitations with energy-gap, $J$. The quintets, that cost an energy $3J$ from the ground state, are the next higher excitations. Since they are safely above the triplets,  
in the simplest approximation, we ignore the quintets and the highest energy heptets in writing a low-energy theory of the trimerized KHA model, $\Hhat$. Thus, we restrict the triangle's Hilbert space to have the singlet, $|s\rangle$, and all the 9 triplets, $|t_{m\nu}\rangle$, only. This reduced problem would nevertheless be sufficient to do a basic stability check of the TS ground state. 

Like the bond-operator representation, that is known to be so useful to the studies of 
dimer phases~\cite{Sachdev1990,Kumar2008,Kumar2010}, we derive here a plaquette-operator representation for the spin-1 operators on a triangular plaquette in the reduced basis, $\{|s\rangle, |t_{m\nu}\rangle\}$. For this, let us first introduce the singlet and triplet plaquette-operators, $\shat^\dag$ and $\that^\dag_{m\nu}$, that are defined as follows.
\begin{subequations}
\begin{align}
|s\rangle&:= \shat^\dagger |\mbox{\o}\rangle \\
|t_{m\nu}\rangle &:= \that_{m\nu}^\dagger |\mbox{\o}\rangle \label{eq:tri_op}
\end{align}
\end{subequations}
Here, $\shat^\dag$ are $t^\dag_{m\nu}$ are the bosonic creation operators in a Fock space with vacuum, $|\mbox{\o}\rangle$. 
The projection of  the infinite dimensional Fock space onto the 10-dimensional Hilbert space spanned by $|s\rangle$ and $|t_{m\nu}\rangle$ is done by the following constraint on the number of these bosons.
\begin{equation}
\shat^\dag\shat + \sum_{m,\nu} \that^\dag_{m\nu}\that^{ }_{m\nu} = 1
\label{eq:constraint}
\end{equation}
In terms of the singlet and triplet plaquette-operators introduced above, the Hamiltonian of a triangle in the reduced basis can be written as follows.  
\begin{equation}
\Hhat_\triangle \approx  -3J\shat^\dag\shat -2J\sum_{m,\nu}\that^\dag_{m\nu}\that^{ }_{m\nu}
\label{eq:Hlocal}
\end{equation}

Next we represent the spin-1 operators of a triangle in terms of the plaquette-operators. Below they are written in an approximate form that is simple and useful. For more details, please refer to Appendix~\ref{sec:rep}.
\begin{subequations}
\label{eq:rep}
\begin{align}
S_{j,z} &\approx \frac{\sbar}{\sqrt{3}} \left\{ c_j\Qhat_{z\onebar} - s_j\Qhat_{z1} \right\} \label{eq:rep-z}\\
S_{j,\alpha} &\approx \frac{2\sbar}{\sqrt{3}} \left\{ c_{j-1} \Qhat_{\alpha\onebar}  -s_{j-1}\Qhat_{\alpha1} \right\}~\mbox{for}~\alpha=x,y \label{eq:rep-xy}
\end{align}
\end{subequations}
Here, $j=1, 2, 3$ denote the spins of a triangle (see Fig.~\ref{fig:TKagome} for spin labels), and $\alpha=x,y$ and $z$ their components. Moreover, $c_j=\cos{\left(\frac{2\pi j}{3}\right)}$ and $s_j=\sin{\left(\frac{2\pi j}{3}\right)}$. The `coordinate' operators $\Qhat_{\alpha\nu}$  for $\nu=1,\onebar$ and $\alpha=x,y$ and $z$ are defined as: $\Qhat_{\alpha\nu} = (\that^\dag_{\alpha\nu}+\that^{ }_{\alpha\nu})/\sqrt{2}$, where the operators, $\that_{\alpha\nu}$, are given as follows~\footnote{The  triplon operators, $\that_{m\nu}$, of Eqs.~(\ref{eq:talphanu}) are obtained from the original triplon operators defined in Eq.~(\ref{eq:tri_op}) via the following simple rotation:
\begin{eqnarray*}
\that_{m1}~ \mbox{\tiny [of Eq.~\eqref{eq:talphanu}]} &=& \frac{1}{\sqrt{2}}\left(\that_{m\onebar}+\that_{m1}\right)~ \mbox{\tiny [of Eq.~\eqref{eq:tri_op}]} \\
\that_{m\onebar}~ \mbox{\tiny [of Eq.~\eqref{eq:talphanu}]} &=& \frac{1}{\sqrt{2}}\left(\that_{m\onebar}-\that_{m1}\right)~ \mbox{\tiny [of Eq.~\eqref{eq:tri_op}]} 
\end{eqnarray*}
}.
\begin{subequations}
\label{eq:talphanu}
\begin{align}
\that_{z\nu} &= (-i)^{\frac{1-\nu}{2}} \that_{0\nu} \\
\that_{x\nu} &= (-i)^{\frac{1-\nu}{2}}(\that_{\bar{1}\nu} -\that_{1\nu})/\sqrt{2} \\
\that_{y\nu} &=i^{\frac{1+\nu}{2}} (\that_{\bar{1}\nu} + \that^{ }_{1\nu})/\sqrt{2}
\end{align}
\end{subequations}
Likewise, we define the conjugate `momentum' operators, $\Phat_{\alpha\nu} = i(\that^\dag_{\alpha\nu}-\that^{ }_{\alpha\nu})/\sqrt{2}$, such that $[\Qhat_{\alpha\nu},\Phat_{\alpha^\prime\nu^\prime}]=i\delta_{\alpha\alpha^\prime}\delta_{\nu\nu^\prime}$ and $\Phat^2_{\alpha\nu}+\Qhat^2_{\alpha\nu} = 2\that^\dag_{\alpha\nu}\that^{ }_{\alpha\nu} + 1$. This canonical change of variables (from $\that$, $\that^\dag$ to $\Phat$,$\Qhat$) is found to be convenient for further analysis. Since the $\nu=0$ triplet operators, $\that_{m0}$, do not appear in Eqs.~(\ref{eq:rep}), we keep them as they are. 

Apart from neglecting the quintets and heptets, we have made two other simplifying approximations in writing Eqs.~(\ref{eq:rep}). One, we have treated the singlet operator, $\shat$, as {\em mean-field}, $\sbar$. Through $\sbar$, which is a measure of the singlet amplitude per triangle, we describe in mean-field approximation the TS phase on Kagom\'e lattice. Two, we have ignored the terms bilinear in triplet operators (see Appendix~\ref{sec:rep}), which amounts to neglecting the interaction between triplets in the effective theory. These are two basic approximations of the mean-field triplon analysis. For a general discussion on triplon mean-field theory, please take a look at the Refs.~\onlinecite{Sachdev1990,Kumar2008,Kumar2010}.

\subsection{\label{subsec:triplon-mft} Plaquette-Triplon Mean-Field Theory}
Now we turn to the model $\Hhat$ of Sec.~\ref{sec:model},  and work out an effective theory for it in terms of the plaquette-operators introduced in the previous subsection. We rewrite the intra-triangle interactions ($J$ terms) in $\Hhat$ as Eq.~(\ref{eq:Hlocal}), with $\shat$ replaced by the uniform TS mean-field, $\sbar$. We also add to it the local constraint, $\sbar^2 + \sum_{\alpha\nu}\that^\dag_{\alpha\nu}\that_{\alpha\nu} =1 $, through an average Lagrange multiplier, $\lambda$. The inter-triangle interactions ($J^\prime$ and $J^{\prime\prime}$ terms) in $\Hhat$ are rewritten using Eqs.~(\ref{eq:rep}). These steps lead to an effective bilinear problem of triplons that, after Fourier transformation, takes the following form in the momentum space. 
\begin{align}{\label{eq:H-trip}}
 \Hhat_{t}  =~& e_{0}N + \sum_{\k}\sum_{\alpha=x,y,z} \Bigg\{ \lambda \left[ \that^{\dagger}_{\alpha0}(\k) \that_{\alpha0}(\k)+\frac{1}{2} \right] \nonumber \\
& + \frac{1}{2}\left[ \lambda\, {\bf \Phat}^\dag_{\alpha}(\k) {\bf \Phat}_{\alpha}(\k) + {\bf \Qhat}^\dag_{\alpha}(\k)\,\mathcal{V}_{\alpha,\k}\,{\bf \Qhat}_{\alpha}(\k) \right] \Bigg\} 
\end{align}
Here, $N$ is the total number of triangular unit-cells in the Kagom\'e lattice, and $e_{0}=-\bar{s}^{2}J+\lambda\bar{s}^{2}-2J-\frac{11}{2}\lambda$. 
Moreover, the operators
 \begin{equation}
 {\bf \Qhat}_\alpha(\k) = \left[\begin{array}{c} \Qhat_{\alpha1}(\k) \\ \Qhat_{\alpha\onebar}(\k) \end{array}\right]~~\mbox{and}~~{\bf \Phat}_\alpha(\k) = \left[\begin{array}{c} \Phat_{\alpha1}(\k) \\ \Phat_{\alpha\onebar}(\k) \end{array}\right],
 \end{equation}
where $\Qhat_{\alpha1}(\k)$ and $\Qhat_{\alpha\onebar}(\k)$ are the Fourier components of $\Qhat_{\alpha 1}(\r)$ and $\Qhat_{\alpha\onebar}(\r)$, respectively. That is, $\Qhat_{\alpha\nu}(\r) = \frac{1}{\sqrt{N}}\sum_\k e^{i\k\cdot\r}\Qhat_{\alpha\nu}(\k)$ for $\nu=1,\onebar$. Here, $\r$ denotes the position vector of the triangular units of Kagom\'e lattice (see Fig.~\ref{fig:TKagome}), and $\k$ is the wavevector in the first Brillouin zone of the corresponding reciprocal lattice (see Figs.~\ref{fig:dispersion_nn} and~\ref{fig:Sq00}). Likewise, $\Phat_{\alpha\nu}(\r) = \frac{1}{\sqrt{N}}\sum_\k e^{i\k\cdot\r}\Phat_{\alpha\nu}(\k)$. Since $\Qhat_{\alpha\nu}(\r)$ and $\Phat_{\alpha\nu}(\r)$ are Hermitian, therefore, $\Qhat^\dag_{\alpha\nu}(\k) = \Qhat_{\alpha\nu}(-\k)$ and $\Phat^\dag_{\alpha\nu}(\k) = \Phat_{\alpha\nu}(-\k)$. Moreover, $[\Qhat_{\alpha\nu}(\k),\Phat_{\alpha^\prime \nu^\prime}(\k^\prime)] = i\delta_{\alpha\alpha^\prime}\delta_{\nu\nu^\prime}\delta_{\k+\k^\prime=0}$, while the $\Qhat_{\alpha\nu}(\k)$'s commute amongst themselves and the same for $\Phat_{\alpha\nu}(\k)$'s.

Since the $\nu=0$ triplon modes, denoted here by $\that_{\alpha 0}(\k) =\frac{1}{\sqrt{N}}\sum_\r e^{-i\k\cdot\r} \that_{\alpha 0}(\r)$, stay decoupled and local, the effective triplon model, $\Hhat_t$, is essentially a problem of two coupled `oscillators', described by $\Qhat_{\alpha1}(\k)$ and $\Qhat_{\alpha\onebar}(\k)$ that are coupled in Eq.~(\ref{eq:H-trip}) via 
\begin{equation}
\mathcal{V}_{\alpha,\k}=\left[\begin{array}{lcl}  \lambda-2\sbar^2\epsilon_{\alpha1,\k} && \sbar^2\eta_{\alpha,\k} \\ 
\sbar^2\eta^*_{\alpha,\k} && \lambda-2\sbar^2\epsilon_{\alpha\onebar,\k} \end{array}\right].
\label{eq:V}
\end{equation}
The $\mathcal{V}_{\alpha,\k}$ is a Hermitian matrix, with $\eta^*_{\alpha,\k}$ as the complex conjugate of $\eta_{\alpha,\k}$. The $\epsilon_{\alpha\nu,\k}$ and $\eta_{\alpha,\k}$ are given below.
\begin{subequations}
\begin{eqnarray}
 \epsilon_{x\onebar,\k} &=&\epsilon_{y\onebar,\k} \nonumber \\ 
  &=& \frac{1}{3}\left[ J^\prime \left( 2\cos{\k\cdot\a_{3}}+2\cos{\k\cdot\a_{1}}-\cos{\k\cdot\a_{2}} \right) \right. \nonumber \\ 
 & & \left. + J^{\prime\prime}\left( 4\cos{\k\cdot\a_2}+\cos{\k\cdot\a_3}+ \cos{\k\cdot\a_1} \right) \right] \\ 
 \epsilon_{z\onebar,\k} & = & \frac{1}{12} \left[ J^\prime \left(2\cos{\k\cdot\a_{2}}+2\cos{\k\cdot\a_{3}}-\cos{\k\cdot\a_{1}}\right) \right.  \nonumber \\ & &  \left. +J^{\prime\prime}\left(4\cos{\k\cdot\a_{1}}+\cos{\k\cdot\a_{2}}+\cos{\k\cdot\a_{3}}\right) \right]
  \end{eqnarray}
\end{subequations}
\begin{subequations}
\begin{align}
\epsilon_{x1,\k} & = \epsilon_{y1,\k} \nonumber \\
 & =J^\prime \cos{\k\cdot\a_{2}} + J^{\prime\prime}\left(\cos{\k\cdot\a_{3}}+\cos{\k\cdot\a_{1}}\right) \\ 
 \epsilon_{z1,\k} &= \frac{1}{4} \left[J^\prime  \cos{\k\cdot\a_{1}} + J^{\prime\prime}\left(\cos{\k\cdot\a_{2}}+\cos{\k\cdot\a_{3}}\right) \right]
  \end{align}
 \end{subequations}
 \begin{subequations}
\begin{eqnarray}
 \eta_{x,\k} &=& \eta_{y,\k} \nonumber \\
 & =& \frac{2}{\sqrt{3}} \Big\{ J^\prime\left[ e^{i\k\cdot\a_3}-e^{i\k\cdot\a_1}-i\sin{\k\cdot\a_2} \right] \nonumber \\
 & & + J^{\prime\prime} \big[ i\left(\sin{\k\cdot\a_1} + 2\sin{\k\cdot\a_2} -\sin{\k\cdot\a_3} \right)  \nonumber \\
 & & + e^{i\k\cdot\a_1} -e^{i\k\cdot\a_3} \big] \Big\} \\
 \eta_{z,\k} & = & \frac{1}{2\sqrt{3}} \Big\{ J^\prime \left[e^{-i\k\cdot\a_2}-e^{-i\k\cdot\a_3}-i\sin{\k\cdot\a_1}\right] \nonumber \\ && + J^{\prime\prime} \big[ i\left(2\sin{\k\cdot\a_1} + \sin{\k\cdot\a_2}-\sin{\k\cdot\a_3} \right) \nonumber \\ 
 && -e^{-i\k\cdot\a_2}+e^{-i\k\cdot\a_3}\big] \Big\}
\end{eqnarray}
 \end{subequations}
Here, $\a_1$ and $\a_2$ are the primitive vectors of the trimerized Kagom\'e lattice (as shown in Fig.~\ref{fig:TKagome}), and $\a_3 = \a_1+\a_2$.
 
 The coupled oscillator problem of $\Hhat_t$ can be diagonalized by making a unitary rotation of $\Qhat_{\alpha 1}(\k)$ and $\Qhat_{\alpha\onebar}(\k)$ to the new `coordinates', $\Qhat_{\alpha +}(\k)$ and $\Qhat_{\alpha-}(\k)$, given by
 \begin{equation}
 \left[\begin{array}{c} \Qhat_{\alpha+}(\k) \\ \Qhat_{\alpha-}(\k) \end{array}\right] = \mathcal{U}^\dag_{\alpha,\k} \left[\begin{array}{c} \Qhat_{\alpha1}(\k) \\ \Qhat_{\alpha\onebar}(\k) \end{array}\right]. \label{eq:UQ}
 \end{equation}
The unitary matrix, $\mathcal{U}_{\alpha,\k}$, that diagonalizes $\Hhat_t$
is given as:
\begin{equation}
\mathcal{U}_{\alpha,\k}=\left[\begin{array}{lcr} \cos{\frac{\theta_{\alpha,\k}}{2}} && - e^{-i\phi_{\alpha,\k}}\sin{\frac{\theta_{\alpha,\k}}{2}} \\ 
 e^{i\phi_{\alpha,\k}}\sin{\frac{\theta_{\alpha,\k}}{2}} && \cos{\frac{\theta_{\alpha,\k}}{2}} \end{array}\right],
\end{equation}
where $\theta_{\alpha,\k} = \tan^{-1}{\{|\eta_{\alpha,\k}|/(\epsilon_{\alpha\onebar,\k}-\epsilon_{\alpha 1,\k})\}}$, and $\eta_{\alpha,\k} = |\eta_{\alpha,\k}| e^{-i\phi_{\alpha,\k}}$ with $|\eta_{\alpha,-\k}|=|\eta_{\alpha,\k}|$ and $\phi_{\alpha,-\k} = -\phi_{\alpha,\k}$. 

In the diagonal form, the $\Hhat_t$ can be written as follows.
\begin{align}
 \Hhat_t =&~ e_{0}N + \sum_{\k}\sum_{\alpha=x,y,z} \Bigg\{\lambda \left[ \that^{\dagger}_{\alpha0}(\k) \that_{\alpha0}(\k)+\frac{1}{2} \right] \nonumber \\
&~ + \sum_{\mu=\pm}E_{\alpha\mu,\k}\left[ \that^{\dagger}_{\alpha\mu}(\k) \that_{\alpha\mu}(\k)+\frac{1}{2}  \right] \Bigg\}
\label{eq:Ht-diagonal}
\end{align}
Here, $ \that_{\alpha\mu}(\k) = \sqrt{\frac{E_{\alpha\mu,\k}}{2\lambda}}\Qhat_{\alpha\mu}(\k) + i\sqrt{\frac{\lambda}{2E_{\alpha\mu,\k}}}\Phat_{\alpha\mu}(\k) $ are the renormalized triplon operators, and 
\begin{equation}
E_{\alpha\mu,\k}=\sqrt{\lambda(\lambda-2\sbar^2\xi_{\alpha\mu,\k})} \label{eq:Ek}
\end{equation}
are the triplon energy dispersions with $ \xi_{\alpha\mu,\k} = [(\epsilon_{\alpha\onebar,\k} + \epsilon_{\alpha1,\k})-\mu\sqrt{(\epsilon_{\alpha\onebar,\k}-\epsilon_{\alpha1,\k})^2+|\eta_{\alpha,\k}|^2}]/2$. The label, $\mu=\pm$, for new operators  defined in Eqs.~(\ref{eq:UQ}), is analogous to, but different from the old label $\nu$. For a stable problem of triplons with positive energy dispersions, the ground state is given by the vacuum of the triplon excitations. Thus, for the $\Hhat_t$ of Eq.~(\ref{eq:Ht-diagonal}), we get the following ground state energy per unit-cell.
\begin{equation}
 e_g = e_{0}+\frac{3\lambda}{2}+\frac{1}{2N}\sum_\k\sum_{\alpha=x,y,z}\sum_{\mu=\pm} E_{\alpha\mu,\k} \label{eq:eg}
\end{equation} 
This $e_g$ is a function of two unknown mean-field parameters, $\lambda$ and $\sbar^2$. We determine them by minimizing $e_g$. 
That is, $\partial_\lambda e_g  =0$ and $\partial_{\sbar^2} e_g =0$, which gives us the following mean-field equations.
\begin{subequations}  \label{eq:sc-gap}
 \begin{align}
  \sbar^2 &= 4-\frac{1}{2N}\sum_{\k}\sum_{\alpha=x,y,z}\sum_{\mu=\pm}\frac{\lambda-\sbar^{2}\xi_{\alpha\mu,\k}}{E_{\alpha\mu,\k}} \label{eq:sbar} \\
   \lambda &=J+\frac{\lambda}{2N}\sum_{\k}\sum_{\alpha=x,y,z}\sum_{\mu=\pm}\frac{\xi_{\alpha\mu,\k}}{E_{\alpha\mu,\k}} \label{eq:lam}
 \end{align}
\end{subequations}
The self-consistent solution of the above mean-field equations gives the physical values of $\lambda$ and $\sbar^2$. 

This formulation offers two distinct physical solutions based on whether the triplon dispersions are \emph{gapped} or \emph{gapless}. 
The $\Hhat_t$ has nine triplon dispersions. The three $\that_{\alpha0,\k}$'s have flat dispersions at $\lambda$. Then, there are six non-trivial $E_{\alpha\mu,\k}$. Note that $E_{x\mu,\k}$ is exactly same as $E_{y\mu,\k}$, but they are different from $E_{z\mu,\k}$.
When the minimum of the lowest of these dispersions in the Brillouin zone is strictly greater than zero, it means there is an energy gap that protects the TS ground state against triplon excitations. We surely expect this to happen when $J^\prime$ and $J^{\prime\prime}$ are near about zero. In this `gapped' TS phase, Eqs.~(\ref{eq:sc-gap}) are applicable in the given form.

However, as the inter-triangle couplings grow stronger, the triplon gap may close at some point $\q$ in the Brillouin zone for strong enough $J^\prime$ or $J^{\prime\prime}$. That is, $E_{\alpha\mu,\q}=0$, for some lower triplon branches. If it happens, then the corresponding, $\k=\q$, terms in Eqs.~(\ref{eq:sc-gap}) will become singular, giving rise to triplon condensation described by the condensate density, $n_c$, a third unknown in the problem. But now we also have a third equation, which is the condition of gaplessness, in addition to Eqs.~(\ref{eq:sc-gap}) that also need to be revised for a non-zero $n_c$. 
From our calculations described in the next section, we either get $E_{\alpha\mu,\q}=0$ at $\q=(0,0)$ for $\alpha=x,y$ and $\mu=\pm$ in a region for $J^{\prime\prime}>0$, or we get $E_{\alpha-,\q}=0$ for $\alpha=x,y$ at $\q=\pi(\xhat+\sqrt{3}\yhat)/3 \equiv (\frac{\pi}{3},\frac{\pi}{\sqrt{3}})$ in another region for $J^{\prime\prime}<0$. The other dispersions are found to be always gapped (see Figs.~\ref{fig:dispersion_nn},~\ref{fig:dispersion_nnn_0} and~\ref{fig:dispersion_nnn_K}). 

The revised equations applicable to the gapless case of $\q=(0,0)$ can be written as follows.
\begin{subequations}{\label{eq:gapless1}}
\begin{align}
\lambda =&~ 2\sbar^2\xi_{\alpha\mu,\q}~~(\mbox{same for}~\alpha=x,y~\mbox{and}~\mu=\pm) \\ 
=&~2\sbar^2(J^\prime+2J^{\prime\prime}) \nonumber \\
\sbar^2 =&~ 4-n_c-\frac{1}{2N}\sum_{\k\neq\q}\sum_{\alpha=x,y}\sum_{\mu=\pm} \frac{\lambda-\sbar^2\xi_{\alpha\mu,\k}}{E_{\alpha\mu,\k}} \nonumber \\
& -\frac{1}{2N}\sum_\k\sum_{\mu=\pm}  \frac{\lambda-\sbar^2\xi_{z\mu,\k}}{E_{z\mu,\k}} \\
n_{c} =&~ \sbar^2\left(1-\frac{J}{\lambda}\right) - \frac{\sbar^2}{2N}\sum_{\k\neq\q}\sum_{\alpha=x,y}\sum_{\mu=\pm} \frac{\xi_{\alpha\mu,\k}}{E_{\alpha\mu,\k}} \nonumber \\
& -\frac{\sbar^2}{2N}\sum_\k\sum_{\mu=\pm}  \frac{\xi_{z\mu,\k}}{E_{z\mu,\k}}
 \end{align}
\end{subequations}
The equation for $\lambda$ here follows directly from the zero gap condition, $E_{\alpha\mu,\q}=0$. The other two equations are derived from Eqs.~(\ref{eq:sc-gap}) by defining the condensate density, $n_c$, as the contribution of the singular terms in Eq.~(\ref{eq:sbar}). That is, $n_c\equiv \frac{1}{2N}\sum_{\alpha=x,y}\sum_{\mu=\pm} (\lambda-\sbar^2\xi_{\alpha\mu,\q})/E_{\alpha\mu,\q}$, in the present case. For the related discussion on triplon analysis, one may look at the Refs.~\onlinecite{Kumar2008,Kumar2010} (which similarly study the dimer problems).

Likewise, in the other gapless phase with $\q=(\frac{\pi}{3},\frac{\pi}{\sqrt{3}})$, the following equations would apply. 
\begin{subequations}{\label{eq:gapless2}}
\begin{align}
\lambda =&~ 2\bar{s}^{2}\xi_{\alpha-,\q}~~(\mbox{same for}~\alpha=x,y) \\
=&~2\sbar^2(J^{\prime}-4J^{\prime\prime}) \nonumber \\
\sbar^2 =&~ 4-n_c-\frac{1}{2N}\sum_{\k\neq\q}\sum_{\alpha=x,y} \frac{\lambda-\sbar^2\xi_{\alpha-,\k}}{E_{\alpha-,\k}} \nonumber \\
& -\frac{1}{2N}\sum_{\k}\left[\sum_{\alpha=x,y} \frac{\lambda-\sbar^2\xi_{\alpha+,\k}}{E_{\alpha+,\k}} + \sum_{\mu=\pm}  \frac{\lambda-\sbar^2\xi_{z\mu,\k}}{E_{z\mu,\k}}\right] \\
n_{c} =&~ \sbar^2\left(1-\frac{J}{\lambda}\right) - \frac{\sbar^2}{2N}\sum_{\k\neq\q}\sum_{\alpha=x,y} \frac{\xi_{\alpha-,\k}}{E_{\alpha-,\k}} \nonumber \\
& -\frac{\sbar^2}{2N}\sum_\k\left[ \sum_{\alpha=x,y} \frac{\xi_{\alpha+,\k}}{E_{\alpha+,\k}} + \sum_{\mu=\pm}  \frac{\xi_{z\mu,\k}}{E_{z\mu,\k}}\right]
 \end{align}
\end{subequations}
Here, $n_c\equiv \frac{1}{2N}\sum_{\alpha=x,y} (\lambda-\sbar^2\xi_{\alpha -,\q})/E_{\alpha -,\q}$.
Physically, a non-zero $n_c$ and a $\q$ account for the AF order with ordering wavevector $\q$ in the ground state. Equations~(\ref{eq:mj}) and~(\ref{eq:m-phi}) in Sec.~\ref{subsec:ordered-phases} describe the magnetic moments in terms of $n_c$ and $\q$ in the two ordered phases. 

\section{\label{sec:discussion} Results and discussion}
To determine the ground state properties of $\Hhat$ within the triplon mean-field theory, we numerically solve the self-consistent equations derived in the previous section. In our calculations, we put $J=1$ and take $0\le J^\prime \le 1$. We keep the second-neighbor coupling, $J^{\prime\prime}$, small, but allow it to take both positive and negative values ($|J^{\prime\prime}| \lesssim 0.6$).

\subsection{\label{subsec:gapped-phase} Gapped Trimerized Singlet Phase}
According to the theory presented in the last section, the energy gap to triplon excitations decides if the ground state is non-magnetic (TS) or magnetically ordered. As the trivial case of independent triangles is surely gapped and non-magnetic, a region around $(J^\prime,J^{\prime\prime})=(0,0)$ is also expected to be so. We identify this region of gapped TS phase by following the change in the triplon gap, $\Delta_t$, by gradually increasing the inter-triangle couplings, $J^\prime$ and $J^{\prime\prime}$. If and when the gap closes, that is $\Delta_t$ becomes zero, it marks the quantum phase transition to an AF ordered phase. In the following, we discuss this first in the nearest-neighbor interaction model for $J^{\prime\prime}=0$, and then in the full model including $J^{\prime\prime}$ .
\subsubsection{\label{subsec:nn} $J^{\prime\prime}=0$ }
\begin{figure}[b]
 \centering
 \includegraphics[width=0.8\columnwidth]{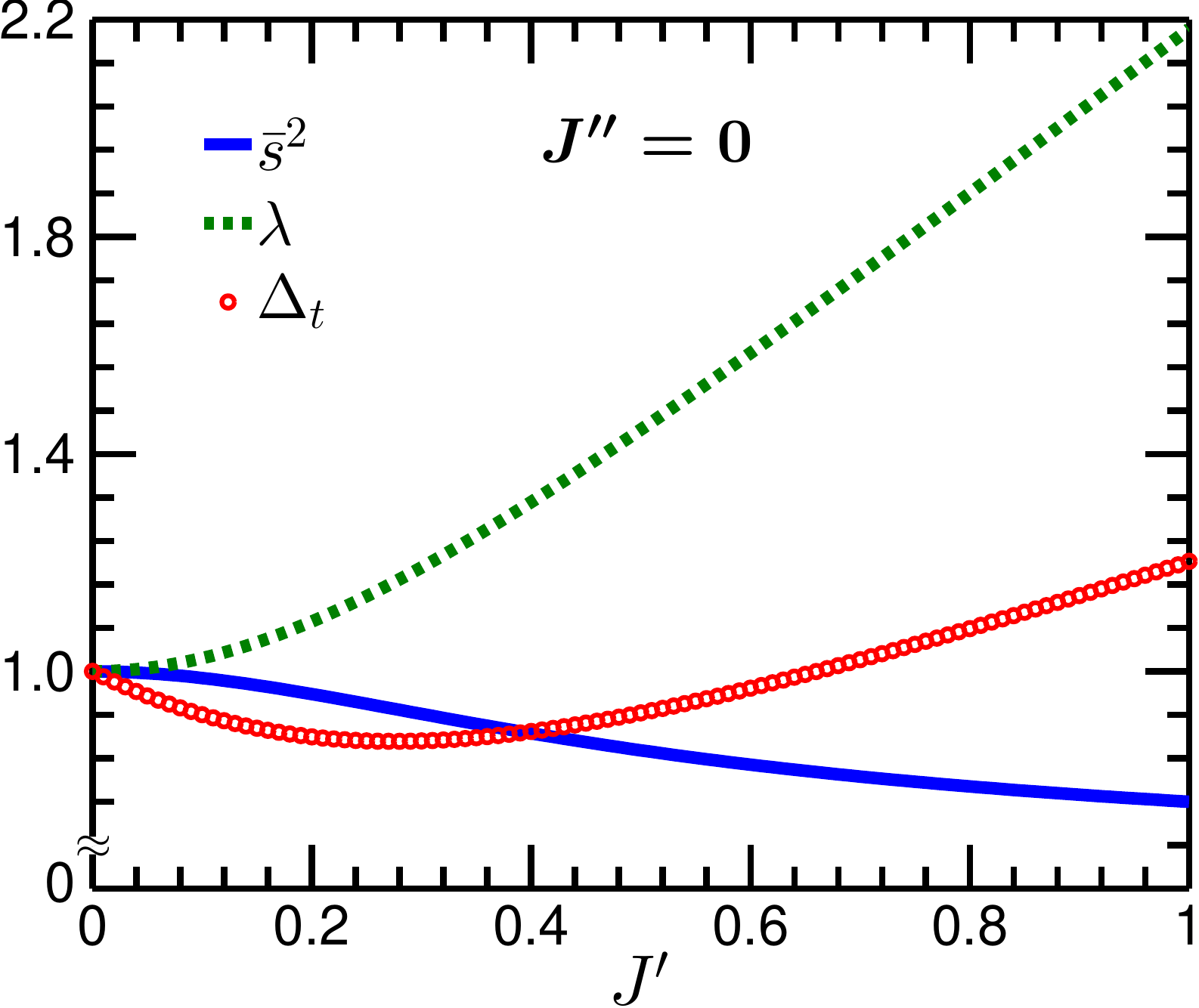}
 \caption{The singlet weight, $\bar{s}^{2}$, the Lagrange multiplier, $\lambda$, and the triplon gap, $\Delta_t$ calculated from the self-consistent Eqs.~(\ref{eq:sc-gap}) for $J^{\prime\prime}=0$ (the nearest-neighbor case of $\Hhat$).}
 \label{fig:gap-Jpp0}
\end{figure}

In this  case, $J^\prime$ is the only interaction variable. We calculate the triplon gap, $\Delta_t$, by solving Eqs.~(\ref{eq:sc-gap}) for $\lambda$ and $\sbar^2$ for different values of $J^\prime$ between 0 and 1. Figure~\ref{fig:gap-Jpp0} presents the calculated values of $\lambda$, $\sbar^2$ and $\Delta_t$ as a function of $J^\prime$. At $J^\prime=0$, it gives $\sbar^2=1$ and $\Delta_t=1$, which is {\em exact} for the independent triangles. A notable feature of this data is the non-zero triplon gap in the entire range of $J^\prime$ between 0 and 1. Although $\Delta_t$ first decreases as $J^\prime$ increases from 0, but then it turns upwards and keeps growing. It is an interesting result which states that, for $J^{\prime\prime}=0$, the non-magnetic TS ground state is stable against triplon excitations, and it \emph{adiabatically} extends all the way upto $J^\prime=1$, starting from the exact case at $J^\prime=0$. This result clearly favors the recent claims of a gapped trimerized singlet ground state for the nearest-neighbor spin-1 KHA model~\cite{Liu2015,Picot2015,Changlani2015}. 
\begin{figure}[htbp]
 \centering
 \includegraphics[width=0.8\columnwidth]{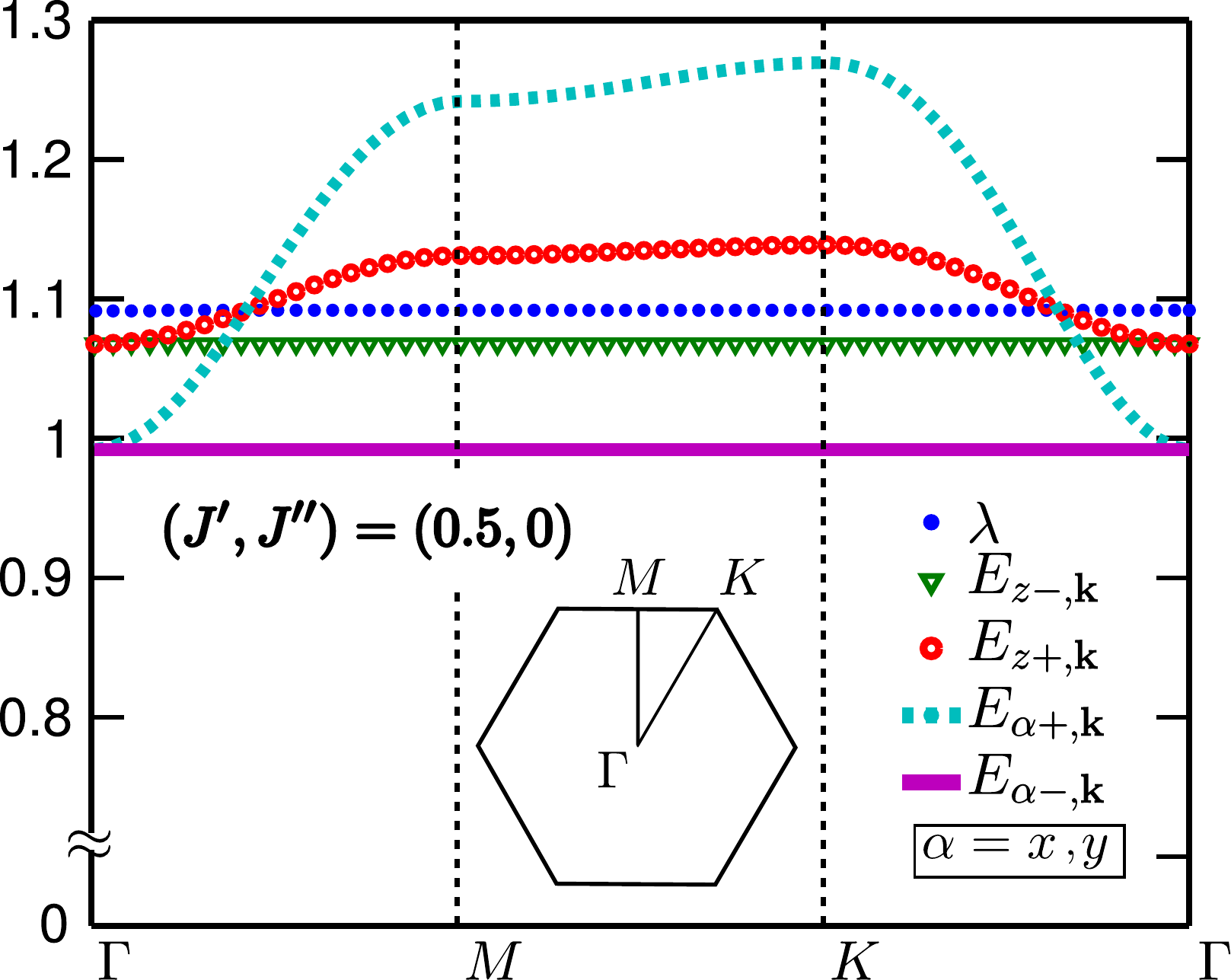}
 \caption{The triplon dispersions [as in Eq.~(\ref{eq:Ht-diagonal})] for the spin-1 trimerized Kagom\'{e} Heisenberg antiferromagnet with $J^{\prime\prime}=0$. Also shown here is the first Brillouin zone of this lattice.}
 \label{fig:dispersion_nn}
\end{figure}

To see where the gap, $\Delta_t$, comes from in the Brillouin zone, we plot the dispersions in Fig.~\ref{fig:dispersion_nn}. Of the nine triplon dispersions given in Eq.~(\ref{eq:Ht-diagonal}), the two flat dispersions, $E_{x-,\k}=E_{y-,\k} = \sqrt{\lambda(\lambda-2\sbar^2J^\prime)}$, are the lowest. Hence, the triplon gap in this case is, $\Delta_t=\sqrt{\lambda(\lambda-2\sbar^2J^\prime)}$.  Two other dispersions, $E_{\alpha+,\k}$ (for $\alpha=x,y$), also become degenerate with the lowest ones at $\k=(0,0)$, the $\Gamma$ point. Moreover, $E_{z-,\k}$ and $\lambda$ (for three $\nu=0$ branches), are also flat. But they are not important for the discussion here, as they are not the lowest in energy. 

\subsubsection{\label{subsec:nnn} $J^{\prime\prime} \neq 0$}
While the non-magnetic TS phase is stable for the nearest-neighbor case of $\Hhat$, it would be nice to know how the second-neighbor interaction, $J^{\prime\prime}$, affects it, or if it generates any magnetic order in the ground state. For the classical KHA problem, it is well-known that even an infinitesimal amount of $J^{\prime\prime}$ causes ordering~\cite{Harris1992}. 

\begin{figure}[htbp]
 \centering
 \includegraphics[width=0.8\columnwidth]{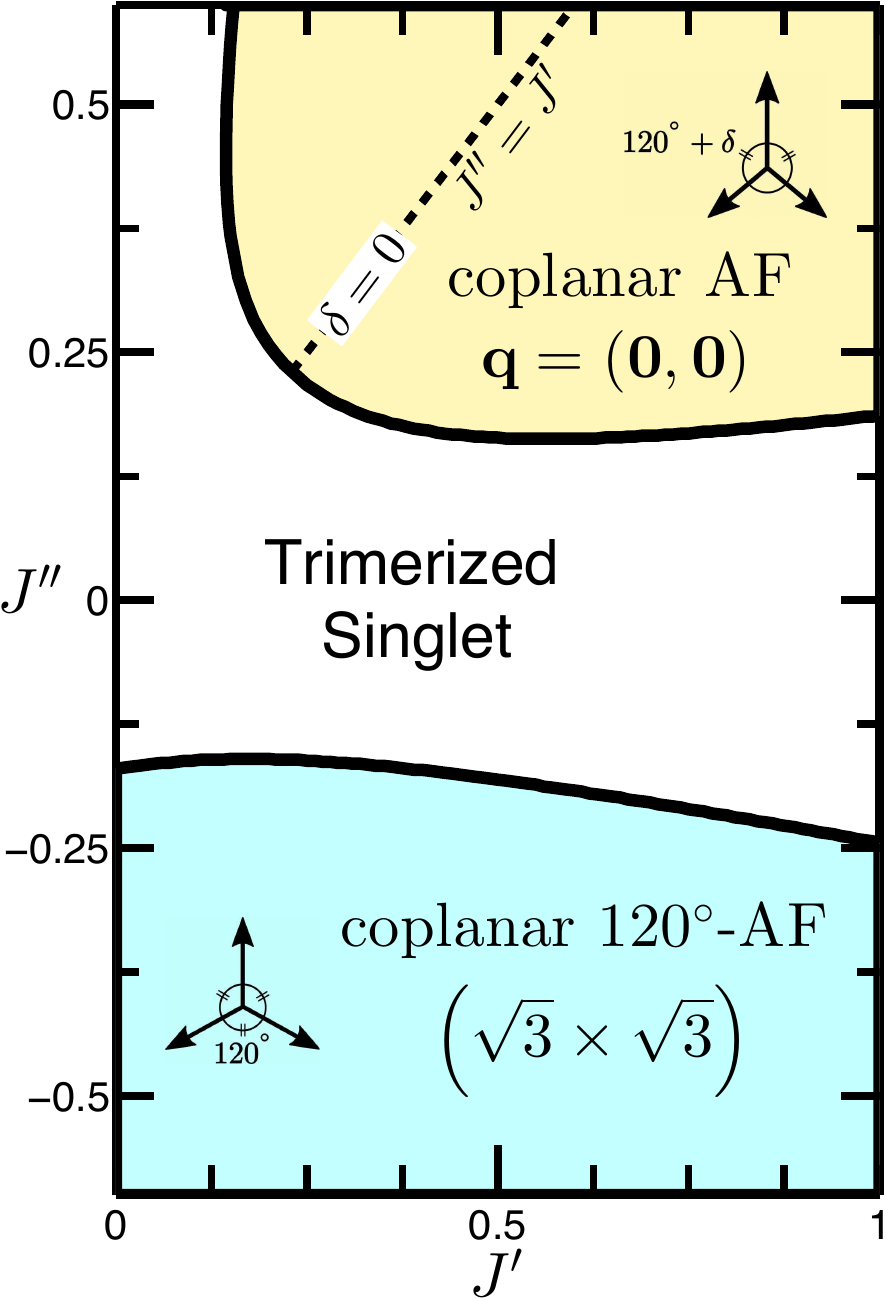}
  \caption{The quantum phase diagram of the spin-1 trimerized Kagom\'e Heisenberg model, $\Hhat$ of Eq.~(\ref{eq:H}), from triplon analysis. The gapped trimerized singlet phase, which is exact at $(J^\prime,J^{\prime\prime})=(0,0)$, extends adiabatically upto $J^\prime=1$, and over a range of $J^{\prime\prime}$. For negative $J^{\prime\prime}$, it undergoes a transition to the gapless phase with coplanar $120^\circ$-antiferromagnetic order having $\sqrt{3}\times\sqrt{3}$ structure. It makes another transition, for positive $J^{\prime\prime}$, to a gapless coplanar antiferromagnetic phase with ordering wavevector, $\q=(0,0)$. In this phase, the magnetic moments deviate from $120^\circ$ orientations by $\delta$, which changes with $J^\prime$ and $J^{\prime\prime}$, and becomes zero only when $J^{\prime\prime}=J^\prime$. }
 \label{fig:QPD}
\end{figure}

For different fixed values of $J^\prime$, we solve Eqs.~(\ref{eq:sc-gap}) with $J^{\prime\prime}$ varying from 0 to $\pm0.6$, and follow the triplon gap. We find that a non-zero $J^{\prime\prime}$ makes the flat modes, $E_{\alpha\mu-,\k}$, dispersive, which reduces the gap, and can even close it altogether. For $J^{\prime\prime}<0$, the triplon gap always closes at some non-zero critical value of $J^{\prime\prime}$. The gap also closes for $J^{\prime\prime} > 0$, but only when $J^\prime > 0.144$. That is, if $J^\prime$ is too small, then the ground state stays gapped for any positive $J^{\prime\prime}$. By scanning the $J^{\prime}$-$J^{\prime\prime}$ plane for the critical points where the triplon gap vanishes, we compute the boundaries of the TS phase. It is found to be stable in an extended region of the $J^\prime$-$J^{\prime\prime}$ plane. For instance, we find the TS phase for $J^\prime=1$ to occur in the range of  $ -0.245 < J^{\prime\prime} < 0.186$, beyond which the triplon gap closes and the AF orders set in. The quantum phase diagram thus generated is shown in Fig.~\ref{fig:QPD}. Clearly, the case of positive $J^{\prime\prime}$ is more frustrated, as it favors the non-magnetic TS phase more than the negative $J^{\prime\prime}$.  

We identify the gapless phase for positive $J^{\prime\prime}$ with wavevector $\q=(0,0)$, the $\Gamma$ point, at which the triplon gap vanishes. In the other gapless phase, for negative $J^{\prime\prime}$, the gap closes at the $K$-point in Brillouin zone, that is $\q=(\pi/3,\pi/\sqrt{3})$. See Figs.~\ref{fig:dispersion_nnn_0} and~\ref{fig:dispersion_nnn_K} for the dispersions in the two phases. As mentioned before, these gapless phases exhibit magnetic order through Bose condensation of triplons with their respective $\q$'s as the ordering wavevectors. The precise forms of the magnetic orders in the two gapless phases are described below. 

\subsection{\label{subsec:ordered-phases} Antiferromagnetically Ordered Phases} 
We now calculate the properties of the gapless phases from Eqs.~(\ref{eq:gapless1}) and~(\ref{eq:gapless2}). These equations enable the computation of triplon condensate density, $n_c$, in addition to giving us the quasiparticle dispersions (and $\sbar^2$). 

The knowledge of $n_c$ is of great physical significance. Together with $\q$, it determines the magnetic order in a gapless phase. 
A non-zero $n_c$ implies spontaneous triplon `displacements', $\langle \Qhat_{\alpha\nu}(\r)\rangle$, which through the plaquette-operator representation given in Eqs.~(\ref{eq:rep}), determine the local magnetic moments, $\langle \Sbf_j(\r) \rangle$, on Kagom\'e lattice. Since the triplons with dispersions $E_{z\mu,\k}$ do not condense (as they are gapped; see Fig.~\ref{fig:dispersion_nnn_0} and~\ref{fig:dispersion_nnn_K}), we get $\langle\Qhat_{z\nu}(\r)\rangle =0$. However, the condensation for $\alpha=x,y$ at $\q$ gives the following non-zero displacements.
\begin{subequations}
\label{eq:Qxy}
\begin{align}
 \langle \Qhat_{x\onebar}\rangle = \sqrt{2n_{c\onebar}}\sin{(\q\cdot\r)},&~ \langle \Qhat_{x1} \rangle = \sqrt{2n_{c1}}\cos{(\q\cdot\r)} \\
 \langle \Qhat_{y\onebar} \rangle =  \sqrt{2n_{c\onebar}}\cos{(\q\cdot\r)},&~ \langle \Qhat_{y1} \rangle= - \sqrt{2n_{c1}}\sin{(\q\cdot\r)} 
 \end{align}
 \end{subequations}
Here, $n_{c1}$ and $n_{c\onebar}$ are the condensate densities for $\nu=1,\onebar$ (that are same for $\alpha=x,y$). Hence, $n_c=2(n_{c\onebar}+n_{c1})$. Since $n_{c1}$ and $n_{c\onebar}$ can in general be different, we define a parameter $\zeta= n_{c1}/n_{c\onebar}$. In terms of $n_c$ and $\zeta$, we can write $n_{c\onebar} = \frac{n_c}{2(1+\zeta)}$ and $n_{c1}=\frac{\zeta n_c}{2(1+\zeta)}$.  

By putting these triplon displacements into the plaquette-operator representation for spins, we get the following general form of the magnetic moments. 
\begin{equation}
\m_j(\r) = m_j (\cos{[\varphi_j-\q\cdot\r]},\sin{[\varphi_j -\q\cdot\r]},0)
\label{eq:mj}
\end{equation}
Here, $\m_j(\r)=\langle\Sbf_j(\r)\rangle$ is the magnetic moment due to $j^{th}$ spin in the triangular unit-cell at position $\r$, with three components, $m_{j,x}(\r)=m_j\cos{[\varphi_j-\q\cdot\r]}$, $m_{j,y}(\r)=m_j\sin{[\varphi_j-\q\cdot\r]}$ and $m_{j,z}(\r)=0$. These moments are obviously \emph{coplanar}. Their amplitudes, $m_j$, and the angles, $\varphi_j$, are given below  for $j=1,2$ and $3$.
\begin{subequations}
\label{eq:m-phi}
\begin{align}
(m_1,\varphi_1) &= \left(2\sbar\sqrt{\frac{n_c}{3(1+\zeta)}},~ \frac{\pi}{2}\right) \\
(m_2,\varphi_2) &= \left(m_1 \frac{\sqrt{1+3\zeta}}{2},~ \varphi_1 + \frac{2\pi}{3}+\delta\right) \\
(m_3,\varphi_3) &= \left(m_2,~ \varphi_1 - \frac{2\pi}{3} - \delta\right) 
\end{align}
\end{subequations}
Here, $\delta = \tan^{-1}{[\sqrt{3}(1-\sqrt{\zeta})/(1+3\sqrt{\zeta})]}$. The $\m_j(\r)$'s on every triangle exactly add up to zero, as it should be in an antiferromagnetic phase. From these general considerations, now we turn to the specific cases.  

\subsubsection{\label{subsubsec:Gamma} Coplanar AF order with $\q=(0,0)$}
\begin{figure}[htbp]
\centering
\includegraphics[width=0.8\columnwidth]{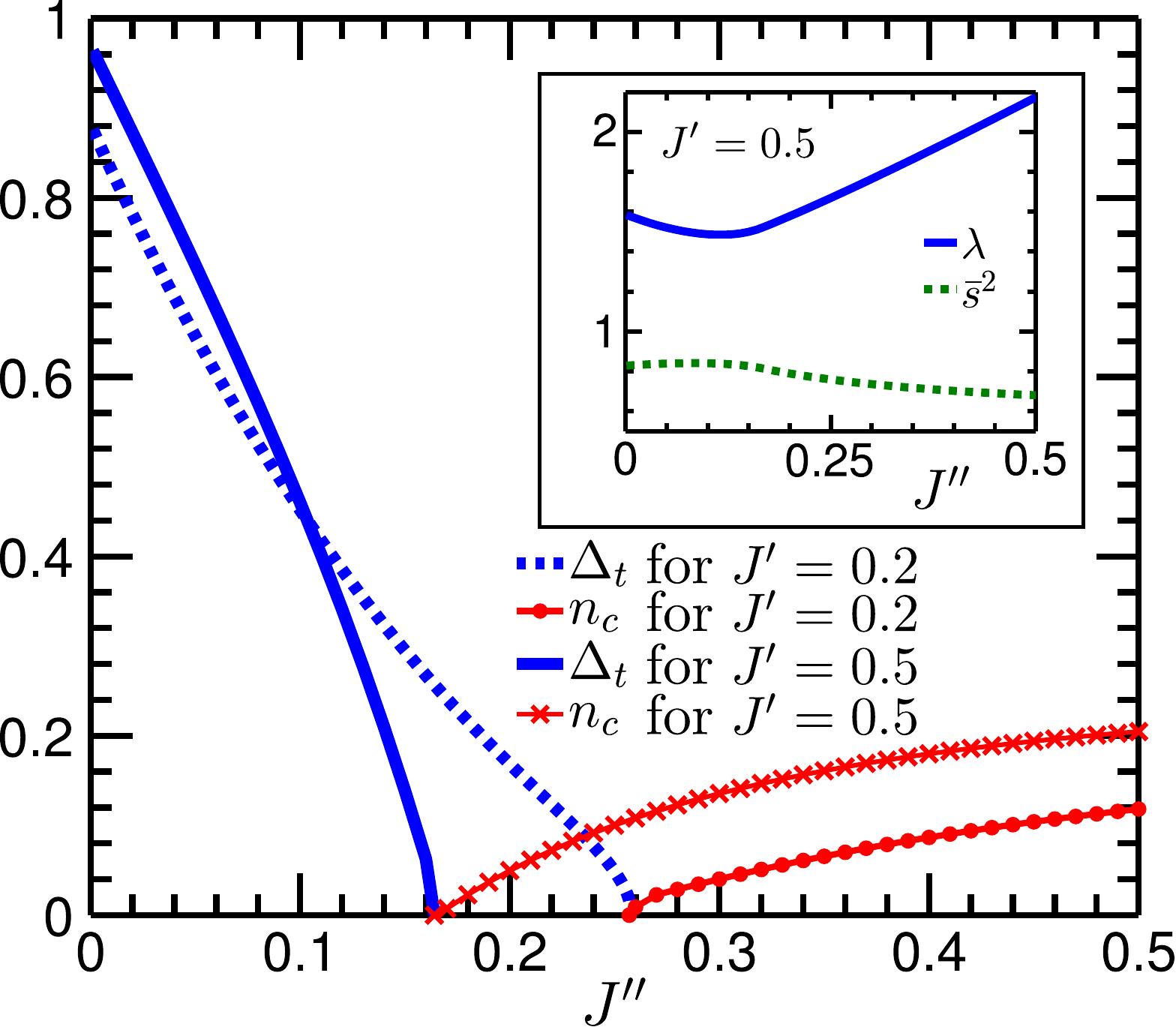}
\caption{The triplon gap, $\Delta_t$, and the condensate density, $n_c$, vs. $J^{\prime\prime}$. Together, they characterize the quantum phase transition from the gapped TS phase to the $\q=(0,0)$ AF ordered phase for $\Hhat$ of Eq.~(\ref{eq:H}). \emph{Inset}: $\lambda$ and $\sbar^2$ vs. $J^{\prime\prime}$.}
\label{fig:gap-nc-G}
\end{figure}

\begin{figure}[b]
 \centering
 \includegraphics[width=0.8\columnwidth]{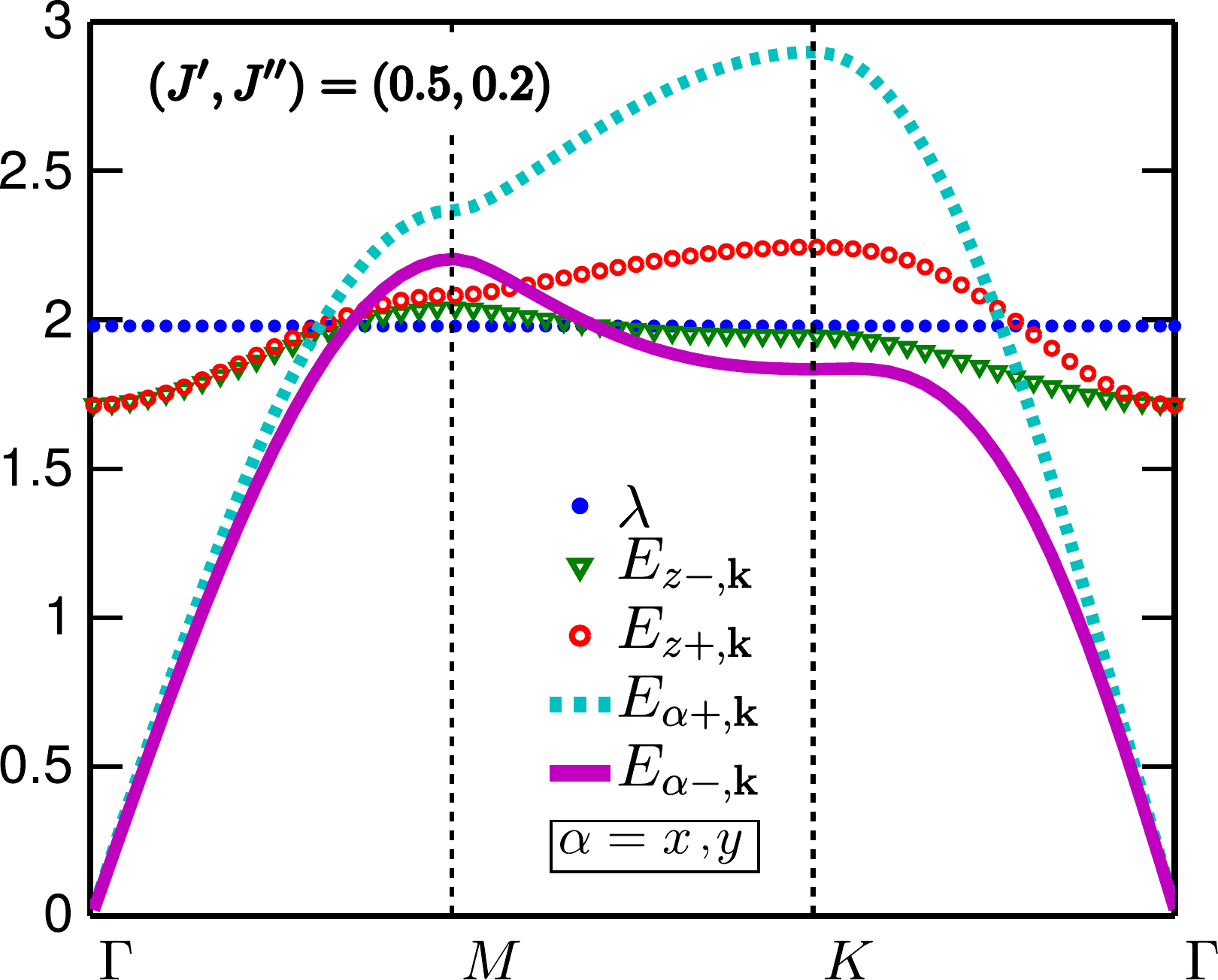}
 \caption{The triplon dispersions [as given in Eq.~(\ref{eq:Ht-diagonal})] in the gapless AF phase with Goldstone mode at $\q=(0,0)$. In this phase, the four dispersions, $E_{\alpha\mu,\k}$ for $\alpha=x,y$ and $\mu=\pm$, go to zero linearly in $|\k|$, at the $\Gamma$ point.}
 \label{fig:dispersion_nnn_0}
\end{figure}

From Eqs.~(\ref{eq:gapless1}), applicable to the phase with Goldstone mode at $\q=(0,0)$, we calculate $\lambda$, $\sbar^2$ and $n_c$. In Fig.~(\ref{fig:gap-nc-G}), we plot $n_c$ as a function of $J^{\prime\prime}$ for fixed values of $J^\prime$, alongside $\Delta_t$ of the gapped phase. It shows a quantum phase transition characterized by the triplon gap that goes to zero continuously at the critical point, and $n_c$ that grows continuously on the other side of the critical point starting from zero at the critical point. The $\sbar^2$ and $\lambda$ also exhibit a kink-like behavior across the transition. In Fig.~\ref{fig:dispersion_nnn_0}, we show the triplon dispersions, of which, the four dispersions with $\alpha=x,y$ and $\mu=\pm$ go to zero linearly in $|\k|$ at $\q=(0,0)$, the $\Gamma$ point. 

\begin{figure}[t]
 \centering
 \includegraphics[width=.8\columnwidth]{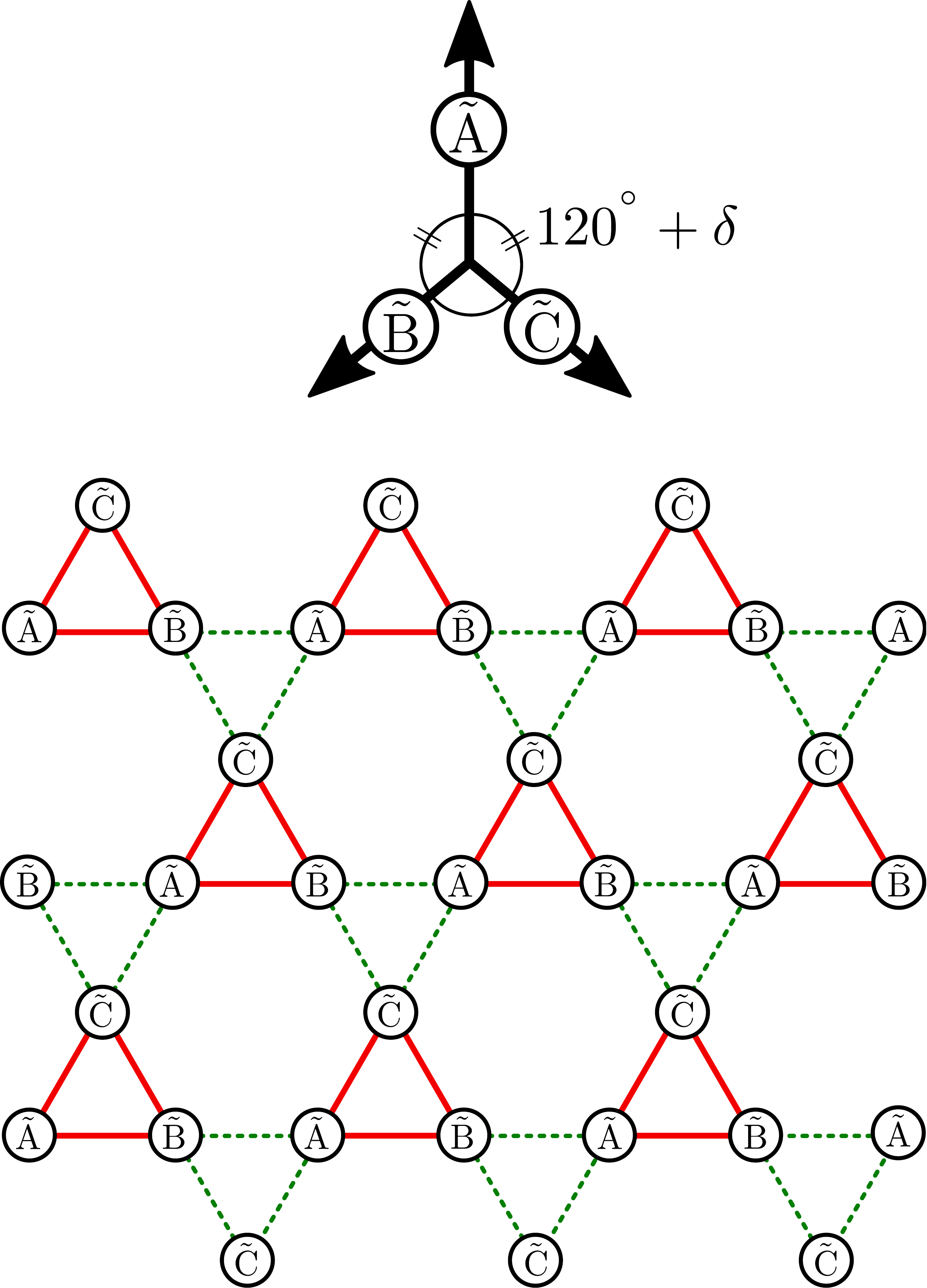}
 \caption{The coplanar antiferromagnetic order with $\q=(0,0)$. Here, the magnetic moments, denoted as $\tilde{A}$, $\tilde{B}$ and $\tilde{C}$, are arranged identically in all the unit-cells (red triangles). The angle between $\tilde{A}$ and $\tilde{B}$ is $120^\circ+\delta$, which is same as the angle between $\tilde{A}$ and $\tilde{C}$.  The magnitude of $\tilde{B}$ is equal to that of $\tilde{C}$, but shorter than that of $\tilde{A}$. Refer to Eqs.~(\ref{eq:m-phi}) for details.}
 \label{fig:order-G}
\end{figure}

We infer the magnetic order in this phase from Eqs.~(\ref{eq:mj}) and (\ref{eq:m-phi}). Its salient features are as follows. Firstly, the magnetic moments are independent of $\r$, obviously because $\q=(0,0)$. That is, the $\m_j(\r)$'s in all the triangular unit-cells look identical, as shown in Fig.~\ref{fig:order-G}. 

\begin{figure}[t]
\centering
\includegraphics[width=.85\columnwidth]{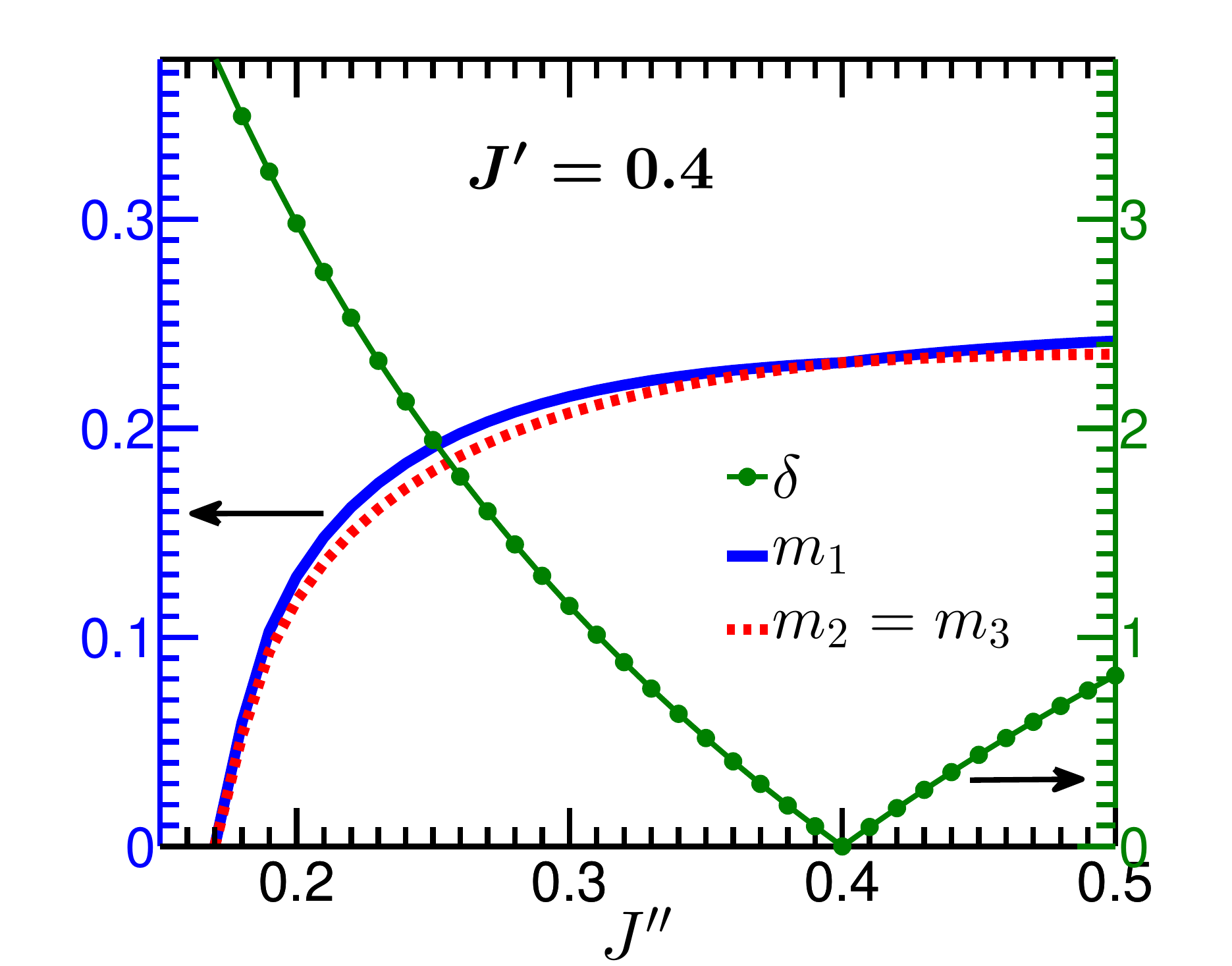}
\caption{The deviation, $\delta$ (in degrees), from the $120^\circ$ orientation and the magnitudes of the magnetic moments ($m_1\ge m_2=m_3$) in the $\q=(0,0)$ phase. Note that $\delta=0$ and $m_1=m_2=m_3$ only when $J^{\prime}=J^{\prime\prime}$.}
\label{fig:delta}
\end{figure}

Secondly, the angles between the magnetic moments in each unit-cell are: $\varphi_{\{2,3\}}-\varphi_1 = 120^\circ+\delta$ and $\varphi_3-\varphi_2=120^\circ-2\delta$, where $\delta$ is non-zero except when $\zeta=1$. Their magnitudes are related as: $m_1 \ge m_2=m_3$, which become equal only when $\zeta=1$. See Fig.~\ref{fig:delta} for typical values of $\delta$ and $m_j$'s which depend upon $J^\prime$ and $J^{\prime\prime}$. These features are clearly {\em at variance} with the perfect $120^\circ$-AF order known from the semiclassical analysis of the KHA problem for large spins~\cite{Harris1992,Chubukov1992}. But then, quite unlike the semiclassical analysis, ours is a calculation with reference to the {\em non-magnetic} TS state, with no presumptions of any magnetic order whatsoever. Here, the magnetic order with a deviation, $\delta$, from the $120^\circ$-AF order has emerged spontaneously through the triplon dynamics present in $\Hhat_t$. For instance, in the other gapless phase for negative $J^{\prime\prime}$ (to be discussed next), the same triplon analysis gives us the perfect $120^\circ$-AF order with $\sqrt{3}\times\sqrt{3}$ structure that is same as known from semiclassical analysis. Hence, the AF order that we  have got here for $\q=(0,0)$ phase looks like a genuine finding. In fact, the semiclassical analysis would miss this order completely, because there the local moments in the reference state are given to be of same magnitudes, which leaves their relative-angles with no choice but to be $120^\circ$ on AF triangles. In our triplon analysis, all of this is decided for itself by the triplon dynamics. We neither fix their magnitudes nor the angles from outside.

This brings us to the third point of note that is about $\zeta$, which effects the deviation of the moments from $120^\circ$ orientation through $\delta$, and makes their magnitudes unequal. The $\zeta$ in this phase arises as the ratio of the slopes (triplon velocities), $v_+$ and $v_-$, of $E_{\alpha+,\k}$ and $E_{\alpha-,\k}$ with respect to $|\k|$ at the $\Gamma$ point. More precisely, $\zeta=\frac{v_-}{v_+}=\sqrt{\frac{J^\prime+2J^{\prime\prime}-|J^\prime-J^{\prime\prime}|}{J^\prime+2J^{\prime\prime}+|J^\prime-J^{\prime\prime}|}}$. As we see in Fig.~\ref{fig:dispersion_nnn_0}, these slopes are visibly different. Therefore, in general, $\zeta<1$ and $\delta\neq 0$. However, when $J^{\prime\prime}=J^\prime$, then $\zeta=1$ and $\delta=0^\circ$. Only in this case, the $\q=(0,0)$ phase has perfect $120^\circ$-AF order of the moments of equal magnitudes. In the quantum phase diagram shown in Fig.~\ref{fig:QPD}, this special case is highlighted by the dashed, $J^{\prime\prime}=J^\prime$, line. On either side of this line, $\zeta<1$ and $\delta\neq 0$. For example, close to the critical point for $J^\prime=1$, $\delta$ is about $9^\circ$. The typical change in $\delta$ and the magnitudes as a function of positive $J^{\prime\prime}$ for a fixed $J^\prime$ is shown in Fig.~\ref{fig:delta}.

\begin{figure}[t]
\centering
\includegraphics[width=.8\columnwidth]{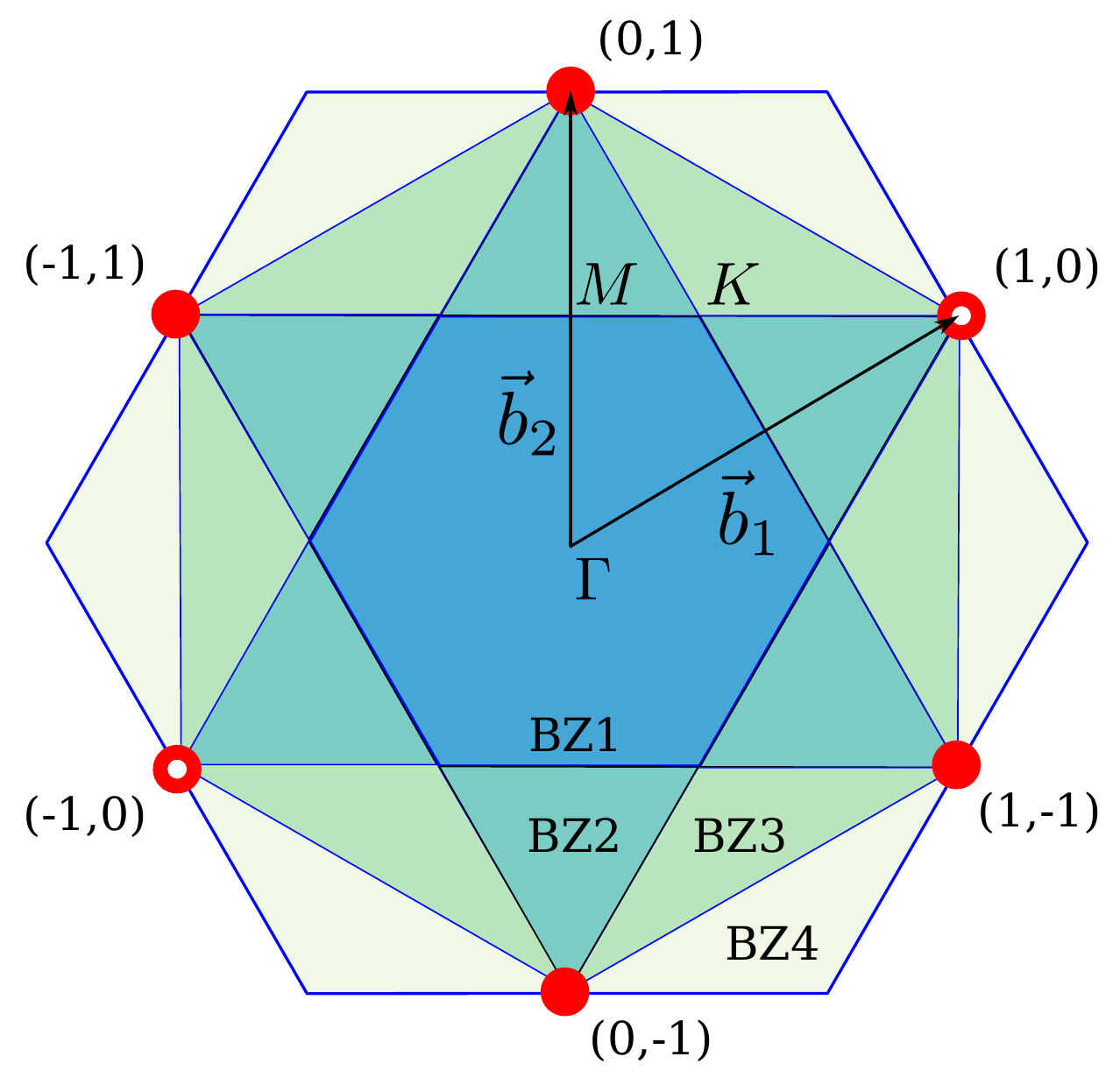}
\caption{The static structure factor, $\mathcal{S}(\k)$, for the coplanar magnetic order (as in Fig.~\ref{fig:order-G}) in the $\q=(0,0)$ phase. Here, $\b_1$ and $\b_2$ are the reciprocal vectors corresponding to $\a_1$ and $\a_2$ of the Kagom\'e lattice (see Fig.~\ref{fig:TKagome}), the points on whose reciprocal lattice are given by $\G=\nu_1\b_1+\nu_2\b_2$, where $\nu_1$ and $\nu_2$ are integers. The four filled red circles at $(\nu_1,\nu_2)= (0,\pm1)$ and $\pm (1,-1)$ denote the Bragg peaks of equal intensity that is different from the intensity of two other equal-intensity Bragg peaks denoted as the hollow red circles at $(\pm 1,0)$. The hollow and the filled circles would all have the same intensity if $\delta=0$. Note that the Bragg peaks occur at some corners of the higher Brillouin zones (BZ2, BZ3, BZ4), but not in the first Brillouin zone (BZ1).}
\label{fig:Sq00}
\end{figure}

Lastly, we discuss an experimental signature of this interesting $\q=(0,0)$ AF order on Kagom\'e lattice. To this end, we calculate the static structure factor of the magnetic moments in this ordered state, which has the translational symmetry of the underlying Kagom\'e lattice, but has the triangular unit-cells of reduced (isosceles as opposed to equilateral) rotational symmetry (due to non-zero $\delta$). We define the static structure factor as:  $\mathcal{S}(\k) \propto |\m (\k)|^2$, where $\m(\k)=\sum_{j, \r} e^{-i\k\cdot\r} \m_j(\r)$, with $\r=l_1\a_1+l_2\a_2$ running over the Bravais lattice of the Kagom\'e lattice ($l_1$ and $l_2$ are integers) and $j=1,2,3$. For the moments, $\m_j(\r)$, given by Eqs.~(\ref{eq:mj}) and~(\ref{eq:m-phi}) for $\q=(0,0)$, we get $\mathcal{S}(\k) \sim f^{ }_\G \, \delta_{\k=\G}$, where $\G=\nu_1\b_1+\nu_2\b_2$ are the points of the reciprocal lattice, and the form factor $f_\G$ is given below.
\begin{align}
\label{eq:fG}
f^{ }_\G &=m_1^2 \left[ (1+3\zeta)\delta_{\nu_2=\mathrm{odd}}+4\delta_{\nu_1=\mathrm{odd}}\,\delta_{\nu_2=\mathrm{even}} \right]
\end{align}
Here, $\nu_1$ and $\nu_2$ are integers, and $\b_1=\pi(\xhat+\frac{1}{\sqrt{3}}\yhat)$ and $\b_2=\frac{2\pi}{\sqrt{3}}\yhat$ are the primitive vectors reciprocal to the $\a_1$ and $\a_2$ of the Kagom\'e lattice (see Figs.~\ref{fig:TKagome} and~\ref{fig:Sq00}). 

The notable features of this $\mathcal{S}(\k)$, that would show up in a neutron diffraction experiment, are as follows. One, the Bragg peaks do not occur when both $\nu_1$ and $\nu_2$ are even integers. That means, no peak at the $\Gamma$ point in the first Brillouin zone (BZ1 of Fig.~\ref{fig:Sq00}). This condition also implies that the Bragg peaks form a Kagom\'e lattice in the reciprocal space. Two, there are two sets of Bragg peaks distinguished by their intensities, $I_{(\nu_1,\nu_2)}$. While the intensities of all the peaks for odd integer values of $\nu_2$ (regardless of $\nu_1$) are same and proportional to $(1+3\zeta)m_1^2 $, for an odd $\nu_1$ and even $\nu_2$, the intensity is proportional to $4m_1^2$. For instance, the Bragg peaks at the four points $(\nu_1,\nu_2)=(0,\pm1)$ and $\pm(1,-1)$ have the same intensity, which is different from the intensity of two other peaks at $(\pm1,0)$. In Fig.~\ref{fig:Sq00}, these two sets of Braggs peaks are shown respectively by the fllled and the hollow red circles. From the ratio of these intensities, one can experimentally measure $\zeta$, and hence $\delta$, the deviation from the $120^\circ$-AF order. One can use the formula
\begin{align}
\label{eq:I10I01}
\frac{I_{(1,0)}}{I_{(0,1)}} &= \frac{4}{1+3\zeta} =  4 \sin^2{\left(\frac{\pi}{6}+\delta\right)} 
\end{align}
to find $\zeta$ and $\delta$. [The $\zeta$, which is the ratio of the triplon velocities at $\Gamma$ point, can also be measured alternatively by measuring triplon dispersions from inelastic neutron scattering.]. Note that, for $\zeta=1$, we get the same intensity for all the Bragg peaks, as it should be for the perfect $120^\circ$-AF order with $\q=(0,0)$ on Kagom\'e lattice~\cite{Messio2011}.

\subsubsection{\label{subsubsec:K} Coplanar $120^\circ$-AF order with $\sqrt{3}\times\sqrt{3}$ structure}
\begin{figure}[t]
\centering
\includegraphics[width=0.8\columnwidth]{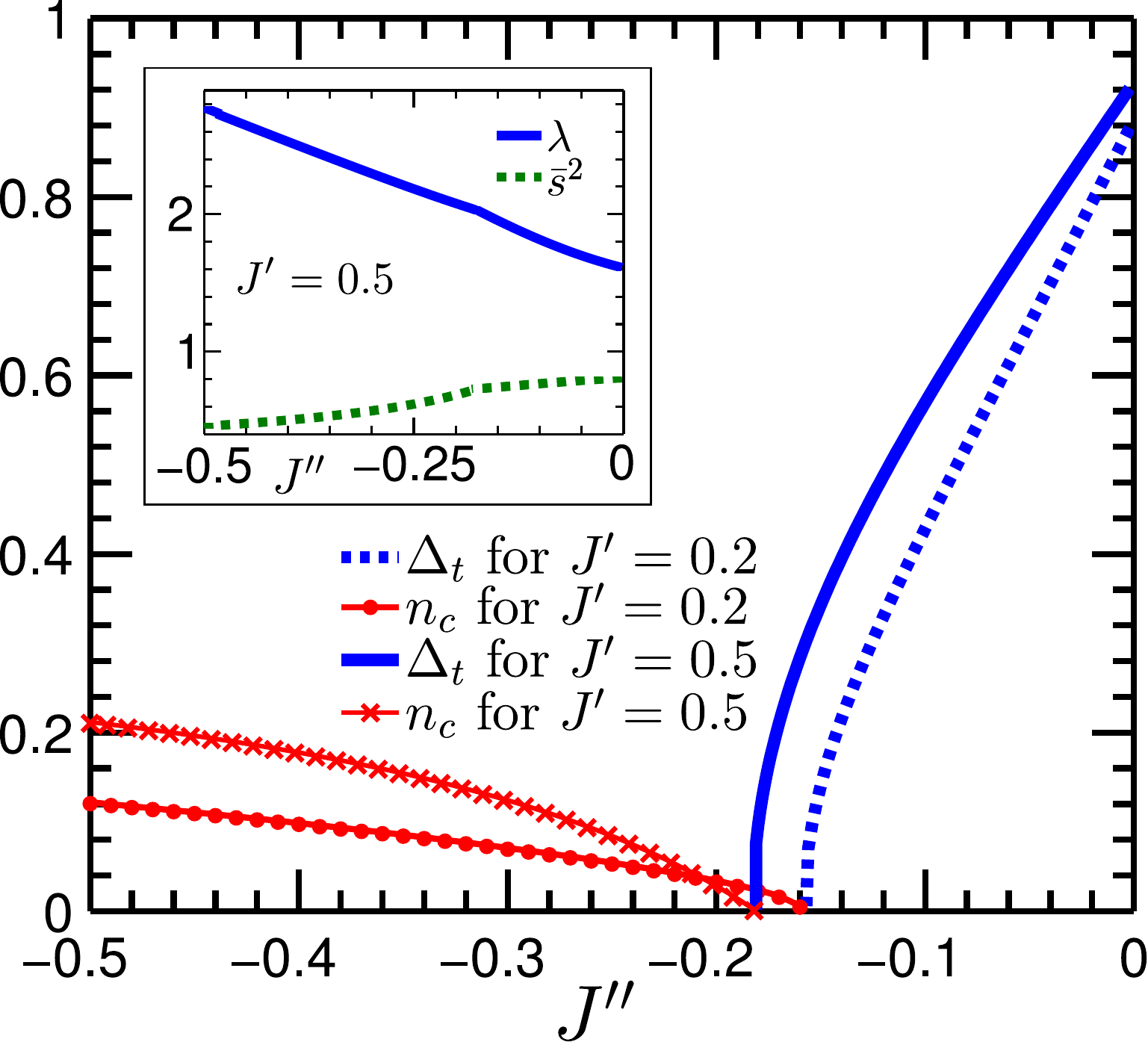}
\caption{The triplon gap, $\Delta_t$, and the condensate density, $n_c$, describing the quantum phase transition from the gapped TS phase to the gapless phase with Goldstone mode at $\q=(\frac{\pi}{3},\frac{\pi}{\sqrt{3}})$. \emph{Inset}: $\lambda$ and $\sbar^2$ vs. $J^{\prime\prime}$.}
\label{fig:gap-nc-K}
\end{figure}

\begin{figure}[t]
 \centering
 \includegraphics[width=0.8\columnwidth]{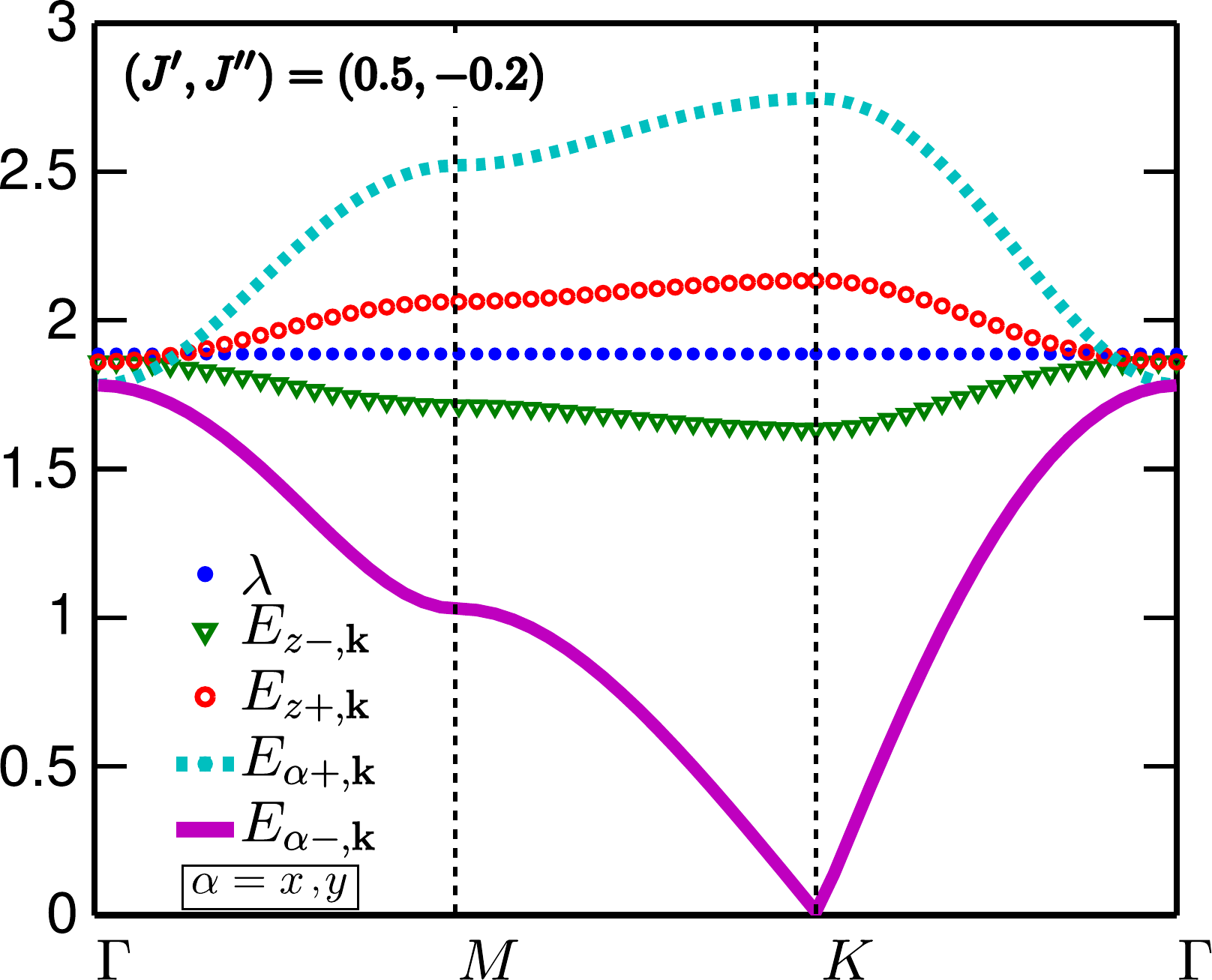}
 \caption{The energy dispersions of the triplon excitations in the gapless AF phase with Goldstone mode at $\q=(\frac{\pi}{3},\frac{\pi}{\sqrt{3}})$. Here, two degenerate dispersions, $E_{x-,\k}$ and $E_{y-,\k}$, go to zero linearly at $\q$, that is, $K$ point  in the Brillouin zone (see Fig.~\ref{fig:dispersion_nn}). }
 \label{fig:dispersion_nnn_K}
\end{figure}

The solutions of Eqs.~(\ref{eq:gapless2}) determine the nature of the gapless phase for negative $J^{\prime\prime}$. The condensate density, $n_c$, and other quantities, calculated as a function of $J^{\prime\prime}$ for fixed $J^\prime$, are plotted in Fig.~\ref{fig:gap-nc-K}. Here again, we see a continuous rise of $n_c$ starting from zero at the critical point.  Moreover, the dispersions, $E_{\alpha-,\k}$ (for $\alpha=x,y$), go to zero linearly at $\q=(\pi/3,\pi/\sqrt{3})=(\b_1+\b_2)/3$, the $K$ point, as shown in Fig.~\ref{fig:dispersion_nnn_K}. 

Interestingly, $\zeta$ is always equal to 1 in this gapless phase. This is so because the condensation of the triplons with dispersion $E_{\alpha-,\k}$ contributes equally to $n_{c1}$ and $n_{c\onebar}$, as $\Qhat_{\alpha-}(\q)$ is an equal weight linear combination of $\Qhat_{\alpha1}(\q)$ and $\Qhat_{\alpha\onebar}(\q)$. Thus, in this phase, we have $\delta=0$, and $m_1=m_2=m_3=\sbar\sqrt{2n_c/3}$. That is, the magnetic moments in every triangle are equal in magnitude, and orientated at $120^\circ$ angle relative to each other. But now they have $\r$-dependent angles, $\varphi_j-\q\cdot\r$, as given in Eq.~(\ref{eq:mj}). It means that the magnetic moments will rotate from one triangle to another, while keeping their internal relative angles fixed at $120^\circ$. 

\begin{figure}[t]
 \centering
 \includegraphics[width=0.9\columnwidth]{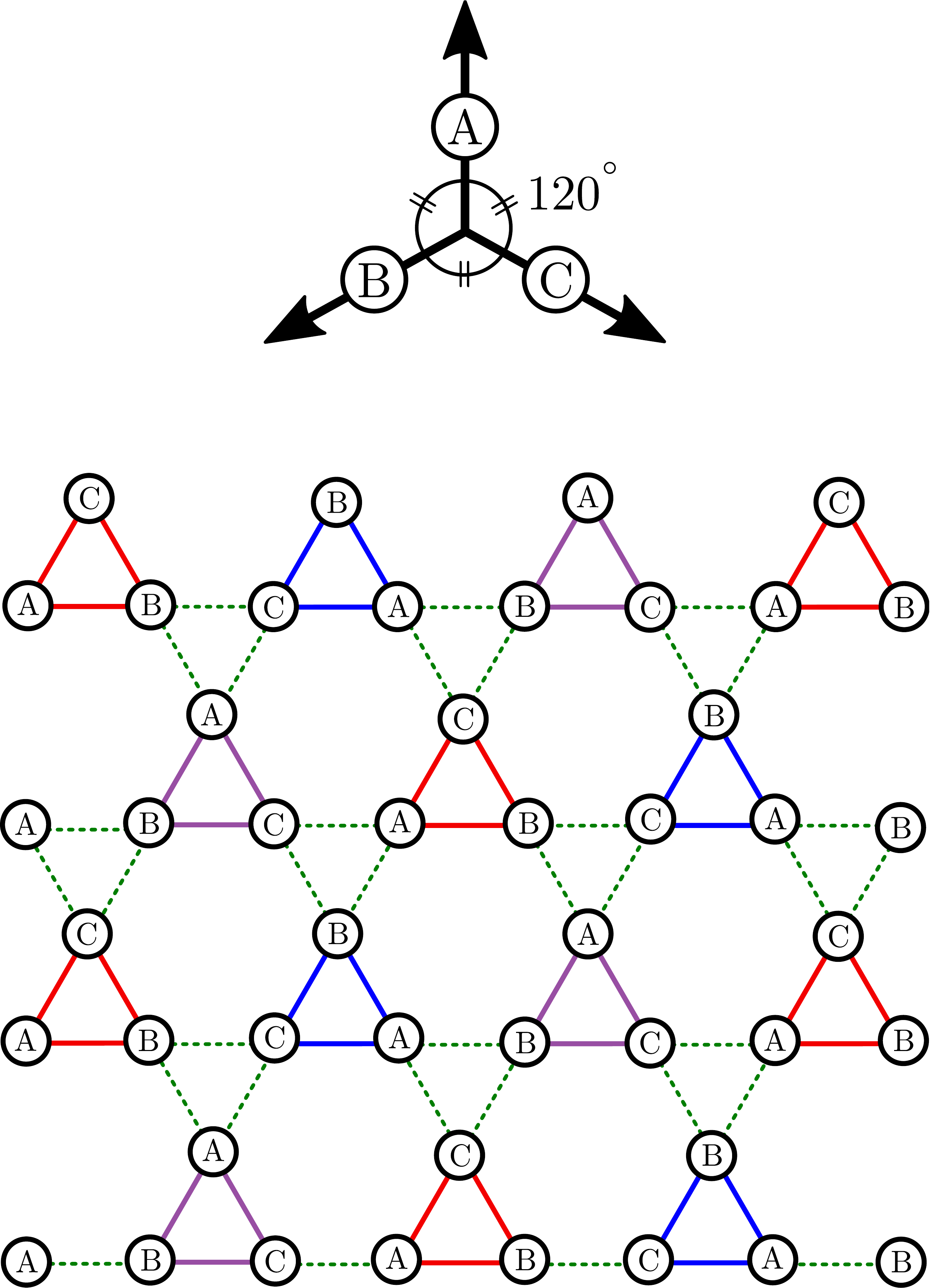}
 \caption{The coplanar $120^\circ$-antiferromagnetic order with ordering wavevector $\q=(\frac{\pi}{3},\frac{\pi}{\sqrt{3}})$ $ =(\b_1+\b_2)/3$. The magnetic moments, denoted as $A$, $B$ and $C$, in every triangle are of equal magnitude, $\sbar\sqrt{2n_c/3}$, and form $120^\circ$ angle relative to each. The moments in the triangles of same color are oriented identically, but rotated by $\pm 2\pi/3$ between the triangles of different colors. The `$\sqrt{3}\times\sqrt{3}$' magnetic lattice, formed of three interpenetrating sublattices of triangles, has a unit-cell consisting of three differently colored triangles.}
 \label{fig:Order_K}
\end{figure}

To understand the $\r$-dependence of the magnetic moments, let us write $\r$ as: $\r=l_1\a_1+l_2\a_2$, where 
$l_1$ and $l_2$ are integers. In the present case, $\q=(\b_1+\b_2)/3$, therefore, $\q\cdot\r = \frac{2\pi}{3}(l_1+l_2)$. It immediately implies that the moments will rotate by $\frac{2\pi}{3}$, if $l_1+l_2$ changes by $-1$ or 2. Or, they will rotate by $-\frac{2\pi}{3}$, if $l_1+l_2$ changes by 1 or $-2$. The  moments will not rotate, however, if the change in $l_1+l_2$ occurs in integer multiples of 3. The magnetic structure that results from these considerations is shown in Fig.~\ref{fig:Order_K}. It consists of three interpenetrating sublattices of the triangles shown in red, blue and purple colors. The magnetic moments in the triangles of one sublattice have same $120^\circ$ orientation, which differs from the orientations on the other two sublattices by $\pm 2\pi/3$. This is the familiar coplanar $120^\circ$-AF order with $\sqrt{3}\times\sqrt{3}$ structure. 
In a diffraction measurement, this magnetic structure would express through the Bragg peaks at $K$-points in the first Brillouin zone, and also at other suitable points in the extended Brillouin zone, as shown in Ref.~\onlinecite{Messio2011}.

We end this section with a brief comparative note on other studies of KHA model of quantum spins with first and second neighbor interactions. As it appears, there are hardly any studies on spin-1 KHA with first and second neighbor interactions, except for a few Schwinger boson calculations which broadly agree on the occurrence of two ordered phases [$\q=(0,0)$ and $\sqrt{3}\times\sqrt{3}$; both with $120^\circ$-AF order] for different signs of $J^{\prime\prime}$ (and $J^\prime=J>0$ in our notation), but undermine the quantum-disordered phase between the two for spin-1~\cite{Motrunich2011,Messio2012}. They do not find any deviation from the $120^\circ$-AF order in the $\q=(0,0)$ phase. The studies on the corresponding spin-1/2 model, that has been investigated more actively by different methods, also claim to find the same ($120^\circ$-AF) ordered phases, separated by a non-magnetic phase~\cite{Suttner2015,Iqbal2015,Kolley2015}. However, in Ref.~\onlinecite{Suttner2015} for spin-1/2 $J_1$-$J_2$ KHA, the $120^\circ$-AF order for $\q=(0,0)$ is noted to be most conspicuous when $J_2 (=J^{\prime\prime}) \approx J_1 (=J^\prime=J)$, and to have enhanced numerical uncertainty away from $J_2/J_1\approx 1$. 
While the first of these observations conforms to having $\delta=0$ on $J^{\prime\prime}=J^\prime$ line in our Fig.~\ref{fig:QPD}, the second one possibly hints at the ordered phase with a non-zero $\delta$. In the light of our Fig.~\ref{fig:Sq00} and Eq.~\ref{eq:I10I01}, a careful relook at the spin structure factor of the spin-1/2 $J_1$-$J_2$ problem would be able to decide if the $\q=(0,0)$ phase there is of the same type as we have found here. In fact, the structure factor calculated in Ref.~\onlinecite{Kolley2015} for $J_2/J_1=0.4$ looks (with naked eyes) very much like our Fig.~\ref{fig:Sq00}, with a slight difference in the intensities at (1,0) and (0,1), and the like points, suggesting a non-zero $\delta$ in the $\q=(0,0)$ phase. 

\section{\label{sec:sum} Summary}
We now conclude by summarizing the main points. Motivated by the current research on spin-1 Kagom\'e quantum antiferromagnets,  we have studied a spin-1 Heisenberg model, the $\Hhat$ of Eq.~(\ref{eq:H}), on trimerized Kagom\'e lattice (see Fig.~\ref{fig:TKagome}). The $\Hhat$ is a problem of coupled antiferromagnetic triangles (with intra-triangle interaction $J=1$), which in the absence of inter-triangle couplings, $J^\prime$ and $J^{\prime\prime}$, trivially realizes the exact TS (trimerized singlet) ground state with zero local magnetic moments and a finite energy gap to triplet excitations. Here, we have studied the stability of this TS ground state, and its transition to ordered phases, as a function of $J^\prime$ and $J^{\prime\prime}$. This we have done by deriving a bosonic plaquette-operator representation for spin-1 operators in terms of the singlet and triplet eigenstates of a triangle [see Eqs.~(\ref{eq:rep}) and the Appendices], and then writing an effective triplon model, $\Hhat_t$ of Eq.~(\ref{eq:H-trip}), for the $\Hhat$ with reference to the TS state. The notable outcomes of this triplon analysis are as follows. 

For $J^{\prime\prime}=0$, that is in the nearest-neighbor case of $\Hhat$, the TS ground state is found to be always gapped and hence stable against triplon excitations. It smoothly extends right upto $J^\prime=1$ (the un-trimerized model), in agreement with the recent numerical findings of the same in the nearest-neighbor spin-1 Kagom\'e Heisenberg antiferromagnetic model~\cite{Liu2015,Picot2015,Changlani2015}. The TS phase is also found to be stable over a range of $J^{\prime\prime}$, before undergoing transition to two gapless ordered AF (antiferromagnetic) phases, one with ordering wavevector $\q=(0,0)$ for positive $J^{\prime\prime}$, and the other with $\q=(\pi/3,\pi/\sqrt{3})$ for negative $J^{\prime\prime}$. The quantum phase diagram obtained from these calculations is presented in Fig.~\ref{fig:QPD}. The magnetic order in the phase with Goldstone modes at $\q=(\pi/3,\pi/\sqrt{3})$ is the familiar coplanar $120^\circ$-AF order with $\sqrt{3}\times\sqrt{3}$ structure (see Fig.~\ref{fig:Order_K}). In the other AF phase with $\q=(0,0)$, the magnetic moments are coplanar, but of different magnitudes (two short and one long in every triangle) and deviate from $120^\circ$ angle relative to each other. These deviations, characterized by an angle $\delta$, are found to arise from the difference in the triplon velocities at $\q=(0,0)$ [see Figs.~\ref{fig:dispersion_nnn_0} and~\ref{fig:order-G}, and Eqs.~(\ref{eq:Qxy}-\ref{eq:m-phi})], and depend on $J^\prime$ and $J^{\prime\prime}$. Only when $J^{\prime\prime}=J^\prime$, the moments become equal in magnitude and form a perfect $120^\circ$-AF order (with $\delta=0)$. This interesting coplanar AF order with a deviation $\delta$, for positive $J^{\prime\prime}$, is a new find with a scope for further investigations in the Kagom\'e antiferromagnets of low quantum spins. 


\begin{acknowledgments}
P.G. acknowledges CSIR (India) for financial support. B.K. acknowledges the financial support under UPE-II scheme of JNU, and DST-FIST support for the computing facility in SPS.
\end{acknowledgments}

\appendix
\section{\label{sec:triangle} Eigenstates of the spin-1 Heisenberg problem on a triangle}
Here, we compute the eigenstates of the Heisenberg model, given below, of three quantum spin-$1$'s.
\begin{equation}
 \Hhat_{\triangle}= J \left(\vec{S}_{1}\cdot\vec{S}_{2}+\vec{S}_{2}\cdot\vec{S}_{3}+\vec{S}_{3}\cdot\vec{S}_{1}\right)
 \label{eq:H_tri}
\end{equation}
The $\Hhat_\triangle$ has spin-rotation symmetry due to which the total-spin, $\Sbf=\Sbf_1+\Sbf_2+\Sbf_3$, is conserved. Therefore, its eigenstates are the total-spin eigenstates given by the total-spin quantum number, $S=0,1,2, 3$, and the total-$S^z$ with values $m=\pm 3,\pm 2, \pm1, 0$. It also has a discrete threefold rotational symmetry that leads to an additional conserved quantum number, $\nu=\pm 1,0$, describing three discrete rotations, $\omega^{\nu}$, of the triangle. Here, $\omega=e^{i2\pi/3}$ is a cube root of unity. Together, these two symmetries make it possible to exactly determine the eigenstates and eigenvalues of $\Hhat_\triangle$.

We denote the product states of three spin-1's as $|m_1 m_2 m_3\rangle$ in the $S_z$ basis, where $m_{j}=1,0,\onebar$ are the eigenvalues of the spin operators, $S_{j,z}$ for $j=1,2,3$.  Here, $\bar{m}$ denotes `$-m$'. We use this notation for writing negative $m$'s.  First, we sectorize these states according to their total-$S_z$ quantum number, $m=m_1+m_2+m_3$, as given in Table~\ref{tab:m-sectors}. Then, we reorganize the states within each $m$-sector according to the quantum number, $\nu$, of the discrete threefold rotation. The basis states in terms of $m$ and $\nu$ are given in Table~\ref{tab:m-nu-states}. Since $m$  and $\nu$ are conserved with respect to $\Hhat_\triangle$, the states from different $m$-$\nu$ subspaces do not mix under $\Hhat_\triangle$. This greatly reduces the eigenvalue problem. We finally write the $\Hhat_\triangle$ as matrix in each $m$-$\nu$ subspace independently, and solve the corresponding eigenvalue problem. The eigenstates $|S,m;\nu\rangle$ of $\Hhat_\triangle$ thus found are given below.

\begin{table}[t]
\caption{\label{tab:m-sectors} The basis states of $\Hhat_\triangle$ according to their total-$S_z$ quantum number, $m$. The states for negative values of $m$ can be obtained from the positive-$m$ states by doing $S_z$ inversion operation, that is, $1\leftrightarrow\onebar$ and $0\leftrightarrow0$.}
\begin{ruledtabular}
\begin{tabular}{ r c | c}
$m$ &&  $|m_{1}m_{2}m_{3}\rangle$ \\ \hline
$3$ && $|111\rangle $\\
\hline
$2$&&$|110\rangle $, $|101\rangle $, $|011\rangle $\\
\hline
$1$&&$|100\rangle $, $|010\rangle $, $|001\rangle $, $|11\bar{1}\rangle $, $|1\bar{1}1\rangle $, $|\bar{1}11\rangle $\\
\hline
$0$&&$|000\rangle $, $|1\bar{1}0\rangle $, $|10\bar{1}\rangle $, $|01\bar{1}\rangle $, $|\bar{1}10\rangle $, $|\bar{1}01\rangle $,  $|0\bar{1}1\rangle $ 
\end{tabular}
\end{ruledtabular}
\end{table}

\emph{Heptets}: These are unique $S=3$ and $\nu=0$ eigenstates of $\Hhat_\triangle$ with eigenvalue $3J$. 
\begin{subequations} \label{eq:heptets}
\begin{eqnarray}
|3,3;0\rangle &=& |111\rangle \\
|3,2;0\rangle & =& \frac{1}{\sqrt{3}}\left( |110\rangle +  |101\rangle + |011\rangle\right) \\
|3,1;0\rangle & =& \frac{1}{\sqrt{15}}\left[2  \left(|100\rangle + |010\rangle + |001\rangle  \right) \right. \nonumber \\
&& \left. + |11\onebar\rangle + |1\onebar1\rangle + |\onebar11\rangle \right] \\
|3,0;0\rangle & =& \frac{1}{\sqrt{10}}\left[2 |000\rangle + |1\onebar0\rangle + |\onebar01\rangle + |01\onebar\rangle \right. \nonumber \\
&& \left. + |\onebar10\rangle + |10\onebar\rangle   + |0\onebar1\rangle \right]
\end{eqnarray}
\end{subequations}
The negative-$m$ eigenstates, $|S,\bar{m};\nu\rangle$, can be obtained by changing $|m_1m_2m_3\rangle$ to $|\bar{m}_1\bar{m}_2\bar{m}_3\rangle$ in the corresponding positive-$m$ eigenstates, $|S,m;\nu\rangle$. For example, $|3,\bar{3}\rangle = |\onebar\onebar\onebar\rangle$, and likewise for other negative-$m$ eigenstates. 

\emph{Quintets}: These are $S=2$ eigenstates with eigenvalue equals to $0$. Here, we get two different sets of quintets, one each for $\nu=1$ and $\onebar$, as written below.
\begin{subequations} \label{eq:quintets}
\begin{eqnarray}
|2,2;\nu\rangle & =& \frac{1}{\sqrt{3}}\left( |110\rangle + \omega^{\bar{\nu}}|011\rangle +  \omega^{\nu} |101\rangle \right) \\
|2,1;\nu\rangle & =& \frac{1}{\sqrt{6}}\left[|100\rangle + \omega^\nu |010\rangle + \omega^{\bar{\nu}} |001\rangle \right. \nonumber \\
&& \left. - \left(|\onebar11\rangle + \omega^\nu |1\onebar1\rangle + \omega^{\bar{\nu}} |11\onebar\rangle  \right) \right] \\
|2,0;\nu\rangle & =& \frac{1}{\sqrt{6}}\left[ |0\onebar1\rangle+\omega^\nu |10\onebar\rangle + \omega^{\bar{\nu}} |\onebar10\rangle\right. \nonumber \\
&& \left. - \left(|1\onebar0\rangle + \omega^\nu |01\onebar\rangle   + \omega^{\bar{\nu}} |\onebar01\rangle\right) \right]
\end{eqnarray}
\end{subequations}
Here, $\bar{\nu} = -\nu$, and the negative-$m$ states can be obtained by doing the $S_z$-inversion ($1\leftrightarrow\onebar$) of the above states.

\begin{table}[t]
\caption{\label{tab:m-nu-states} The basis states of $\Hhat_\triangle$ in terms of the quantum number, $\nu$, of threefold rotation. To obtain negative-$m$ states, do the $S_z$ inversion operation on the  positive-$m$ states.}
\begin{ruledtabular}
\begin{tabular}{r | r | c}
$m$             & $\nu$ 		& Basis states    \\ \hline
$3$              & $0$    		& $|111\rangle$  \\ \hline
\multirow{3}{*}{$2$} & $0$  & $\frac{1}{\sqrt{3}}\left(|110\rangle +|101\rangle +|011\rangle\right)$  \\ \cline{2-3}
& $1$  & $\frac{1}{\sqrt{3}}\left(|110\rangle +\omega |101\rangle +\omega^2 |011\rangle \right)$ \\ \cline{2-3}
& $-1$ & $\frac{1}{\sqrt{3}}\left(|110\rangle +\omega^2 |101\rangle +\omega |011\rangle \right)$ \\ \hline
\multirow{6}{*}{$1$}  & \multirow{2}{*}{$0$} & $\frac{1}{\sqrt{3}}\left(|100\rangle +|010\rangle +|001\rangle \right)$,  \\ 
&	& $\frac{1}{\sqrt{3}}\left(|11\bar{1}\rangle +|1\bar{1}1\rangle +|\bar{1}11\rangle \right)$ \\ \cline{2-3}
& \multirow{2}{*}{$1$} & $\frac{1}{\sqrt{3}}\left(|100\rangle +\omega |010\rangle +\omega^2|001\rangle \right)$, \\ 
&    & $\frac{1}{\sqrt{3}}\left(|11\bar{1}\rangle +\omega|1\bar{1}1\rangle +\omega^2|\bar{1}11\rangle \right)$ \\ \cline{2-3}
& \multirow{2}{*}{$-1$} & $\frac{1}{\sqrt{3}}\left(|100\rangle +\omega^2|010\rangle +\omega|001\rangle \right)$, \\
&    & $\frac{1}{\sqrt{3}}\left(|11\bar{1}\rangle+\omega^2|1\bar{1}1\rangle +\omega|\bar{1}11\rangle \right)$ \\ \hline
\multirow{6}{*}{$0$} & \multirow{3}{*}{$0$}  & $|000\rangle$, $\frac{1}{\sqrt{3}}\left(|10\bar{1}\rangle +|0\bar{1}1\rangle +|\bar{1}10\rangle \right)$, \\
&    & $\frac{1}{\sqrt{3}}\left(|1\bar{1}0\rangle +|\bar{1}01\rangle +|01\bar{1}\rangle \right)$ \\ \cline{2-3}
& \multirow{2}{*}{$1$} & $\frac{1}{\sqrt{3}}\left(|10\bar{1}\rangle +\omega^2 |0\bar{1}1\rangle +\omega|\bar{1}10\rangle \right)$, \\ 
&    & $\frac{1}{\sqrt{3}}\left(|1\bar{1}0\rangle +\omega^2|\bar{1}01\rangle +\omega|01\bar{1}\rangle \right)$ \\ \cline{2-3}
& \multirow{2}{*}{$-1$} & $\frac{1}{\sqrt{3}}\left(|10\bar{1}\rangle +\omega|0\bar{1}1\rangle +\omega^2|\bar{1}10\rangle \right)$, \\
&   & $\frac{1}{\sqrt{3}}\left(|1\bar{1}0\rangle +\omega|\bar{1}01\rangle +\omega^2|01\bar{1}\rangle \right)$                      
\end{tabular}
\end{ruledtabular}
\end{table}
 
\emph{Triplets}: Next, we have three sets of triplets, one each for $\nu=0, 1$ and $\onebar$, with eigenvalue, $-2J$. Thus, $\Hhat_\triangle$ has 9 degenerate $S=1$ eigenstates. Here, we denote the triplet states $|1,m;\nu\rangle$ as $|t_{m\nu}\rangle$. This slight change in notation is introduced to facilitate a convenient notation for the plaquette-operator representation (in the reduced space of the triplets and the singlet), as used in the main text (see Sec.~\ref{subsec:spinoperator}). These triplets are written as follows.
\begin{subequations}{\label{eq:triplets}}
\begin{align}
 & (\mbox{for}~m=0~\mbox{and}~\nu=0) \nonumber \\
|t_{00}\rangle =&~\frac{1}{\sqrt{15}}\left[-3 |000\rangle + |10\bar{1}\rangle +|0\bar{1}1\rangle+|\bar{1}10\rangle \right. \nonumber \\
& \left. + |1\bar{1}0\rangle +|\bar{1}01\rangle+|01\bar{1}\rangle \right] \\
& \nonumber \\
& (\mbox{for}~m=0~\mbox{and}~\nu=1,\onebar) \nonumber \\
|t_{0\nu}\rangle =&~\frac{1}{\sqrt{6}}\left[ |10\bar{1}\rangle +\omega^{\nu}|\bar{1}10\rangle + \omega^{\bar{\nu}} |0\bar{1}1\rangle \right. \nonumber \\
& \left. +|01\bar{1}\rangle + \omega^{\nu}|\bar{1}01\rangle+ \omega^{\bar{\nu}}|1\bar{1}0\rangle  \right]\\
& \nonumber \\
& (\mbox{for}~ m=1,\onebar~\mbox{and}~\nu=0) \nonumber \\
|t_{m0}\rangle =&~\frac{1}{\sqrt {15}}\left[ |m00\rangle +|0m0\rangle +|00m\rangle \right.  \nonumber \\
& \left. -2\left(|\bar{m}mm\rangle +|m\bar{m}m\rangle + |mm\bar{m}\rangle \right) \right]\\
& \nonumber \\
& (\mbox{for}~m=1,\onebar~\mbox{and}~\nu=1,\onebar) \nonumber \\
|t_{m\nu}\rangle  =&~\frac{1}{\sqrt {6}}\left[ |m00\rangle +\omega^{\nu}|0m0\rangle +\omega^{\bar{\nu}}|00m\rangle  \right.\nonumber\\
& \left. + |\bar{m}mm\rangle +\omega^{\nu}|m\bar{m}m\rangle +\omega^{\bar{\nu}}|mm\bar{m}\rangle \right]
\end{align}
\end{subequations}

\emph{Singlet}: Finally, we write the only singlet eigenstate, that is $|0,0;0\rangle$, of $\Hhat_\triangle$. It has has an eigenvalue of $-3J$. Here, we denote it as $|s\rangle$.
\begin{align}
|s\rangle =&~\frac{1}{\sqrt{6}}\left[ |1\bar{1}0\rangle -|\onebar10\rangle  + |\bar{1}01\rangle - |10\onebar\rangle + |01\bar{1}\rangle  - |0\bar{1}1\rangle \right] \label{eq:singlet}
\end{align}

For an antiferromagnetic $\Hhat_\triangle$, that is $J>0$, the singlet at $-3J$ is the lowest energy eigenstate. The triplets at $-2J$ are the lowest excited states, while the quintets and the heptet sit further up at the higher energies. 

\section{\label{sec:rep} Plaquette-operator representation of the spin-1 operators of a triangle}
We now derive the plaquette-operator representation for the spin-1 operators of an antiferromagnetic triangle in its reduced 10-dimensional basis, $\{|s\rangle, |t_{m\nu}\rangle\}$. Here, $|s\rangle$ is the singlet state and $|t_{m\nu}\rangle$'s are 9 degenerate triplets given in Eq.~(\ref{eq:singlet}) and~(\ref{eq:triplets}), respectively. This is the minimal basis that can be used to describe the low-energy dynamics of the trimerized Kagom\'e model, $\Hhat$ of Eq.~(\ref{eq:H}).

The operators $S_{j,z}$ and $S_{j,+}$ are the z-component and the raising operator, respectively, of the $j^{th}$ spin on a triangle, where $j=1,2,3$ (see Fig.~\ref{fig:TKagome} for spin labels).  
Let us, for convenience, denote the 10 basis states as $|b_l\rangle$, where the integer $l$ runs from 0 to 9. 
More precisely,  
\begin{subequations}
\begin{align}
|b_0\rangle &= |s\rangle,~\mbox{and} \\
|b_{l}\rangle &= |t_{m\nu}\rangle~\mbox{for}~l=3m+\nu+5,
\end{align}
\end{subequations}
where both $m$ and $\nu=\onebar,0,1$. In this notation, we can write the spin operators as: $S_{j,z}=\sum_{l,l^\prime} \mathcal{M}^{l l^\prime}_{j,z} |b_l\rangle\langle b_{l^\prime}| $ and  $S_{j,+}=\sum_{l,l^\prime} \mathcal{M}^{l l^\prime}_{j,+} |b_l\rangle\langle b_{l^\prime}| $, where the matrix elements are defined as $\mathcal{M}^{l l^\prime}_{j,z} = \langle b_l |S_{j,z}|b_{l^\prime}\rangle$ and $\mathcal{M}^{ll^\prime}_{j,+} = \langle b_l |S_{j,+}|b_{l^\prime}\rangle$. 
Next we define the bosonic operators, $\bhat_l^\dag$, such that  
\begin{equation}
|b_l\rangle := \bhat^\dag_l |\mbox{\o}\rangle.
\end{equation} 
These `plaquette-operators' (corresponding to the eigenstates of a triangular plaquette) live in a Fock space with vacuum, $|\mbox{\o}\rangle$, and satisfy the constraint, $\sum_l \bhat^\dag_l\bhat^{ }_l =1$. We finally write the plaquette-operator representation of the spin-1 operators on a triangle as:
\begin{equation}
S_{j,z} = \sum_{l,l^\prime} \mathcal{M}^{l l^\prime}_{j,z} \,  \bhat^\dag_l  \bhat^{ }_{l^\prime} ~\mbox{and} ~ 
S_{j,+} = \sum_{l,l^\prime} \mathcal{M}^{l l^\prime}_{j,+} \,  \bhat^\dag_l \bhat^{ }_{l^\prime} ,
\label{eq:rep-general}
\end{equation}
where the matrices $\mathcal{M}_{j,z}$ and $\mathcal{M}_{j,+}$ are given in Eqs.~(\ref{eq:Mz1})-(\ref{eq:Mplus3}), with $l$ going from 0 for the first row to 9 for the last row, and likewise for the column index $l^\prime$.

The general representation in Eq.~(\ref{eq:rep-general}) is the basis of a more simplified plaquette-operator representation in Eqs.~(\ref{eq:rep}) that we have used for doing triplon analysis in the main text. There, we have approximated $\shat$, that is $\bhat_0$, by a mean-field, $\sbar$, and kept only those triplet terms that are coupled with $\sbar$, neglecting the \emph{triplet-only} terms of Eq.~(\ref{eq:rep-general}). This latter approximation amounts to ignoring triplon-triplon interactions the effective theory, akin to ignoring the interaction between magnons in the linear spin-wave analysis. 

\begin{align} 
& \mathcal{M}^z_{1}=  \nonumber\\
&{\footnotesize \left(
 \begin{smallmatrix}
0 & 0 & 0 & 0 & \frac{\omega-1}{6}& 0 & \frac{\omega^{2}-1}{6}& 0 & 0 & 0 \\
0 & -\frac{1}{3}& -\sqrt{\frac{5}{18}} & \frac{1}{6}& 0& 0& 0& 0 & 0 & 0  \\
0 & -\sqrt{\frac{5}{18}} & -\frac{1}{3} & -\sqrt{\frac{5}{18}} & 0 & 0 & 0 & 0 & 0 & 0\\
0& \frac{1}{6}          & -\sqrt{\frac{5}{18}} & -\frac{1}{3} & 0 & 0 & 0 & 0 & 0   & 0  \\
\frac{\omega^{2}-1}{6} & 0  & 0 & 0 & 0 & \frac{\omega^{2}+1}{\sqrt{10}} & \frac{\omega+1}{2}  & 0 & 0  & 0 \\
0& 0  & 0 & 0 & \frac{\omega+1}{\sqrt{10}} & 0 & \frac{\omega^{2}+1}{\sqrt{10}} & 0 & 0   & 0  \\
\frac{\omega-1}{6}  & 0  & 0   & 0  & \frac{\omega^{2}+1}{2} & \frac{\omega+1}{\sqrt{10}} & 0 & 0  & 0   &0 \\
0& 0  & 0 & 0 & 0  & 0& 0 & \frac{1}{3}& \sqrt{\frac{5}{18}} & -\frac{1}{6} \\
0& 0  & 0 & 0 & 0 & 0& 0& \sqrt{\frac{5}{18}} & \frac{1}{3} & \sqrt{\frac{5}{18}} \\
0& 0  & 0  & 0 & 0 & 0 & 0 & -\frac{1}{6}  & \sqrt{\frac{5}{18}} & \frac{1}{3}        
\end{smallmatrix} \right)} \label{eq:Mz1}\\
%
&  \mathcal{M}^+_{1}= \nonumber \\
& {\footnotesize \left(
\begin{smallmatrix}
0 & -i\sqrt{\frac{2}{3}} & 0 & i\sqrt{\frac{2}{3}} & 0 & 0 & 0 & 0 & 0 & 0 \\
0 & 0 & 0 & 0 & 0 & 0 & 0 & 0 & 0 & 0 \\
0 & 0 & 0 & 0 & 0 & 0 & 0 & 0 & 0 & 0 \\
0 & 0 & 0 & 0 & 0 & 0 & 0 & 0 & 0 & 0 \\
0 & \frac{\omega}{3\sqrt{2}} & \frac{\omega}{3\sqrt{5}} & \frac{\omega}{3\sqrt{2}} & 0 & 0 & 0 & 0 & 0 & 0 \\
0 & -\frac{\sqrt{5}}{3} & -\frac{\sqrt{2}}{3} & -\frac{\sqrt{5}}{3} & 0 & 0 & 0 & 0 & 0 & 0 \\
0 & \frac{\omega^{2}}{3\sqrt{2}} & \frac{\omega^{2}}{3\sqrt{5}} & \frac{\omega^{2}}{3\sqrt{2}} & 0 & 0 & 0 & 0 & 0 & 0 \\
-i\sqrt{\frac{2}{3}} & 0 & 0 & 0 & \frac{\omega^{2}}{3\sqrt{2}} & -\frac{\sqrt{5}}{3} & -\frac{\sqrt{2}\omega}{3} & 0 & 0 & 0 \\
0 & 0 & 0 & 0 & \frac{4\omega^{2}}{3\sqrt{5}} & -\frac{\sqrt{2}}{3} & \frac{4\omega}{3\sqrt{5}} & 0 & 0 & 0 \\
i\sqrt{\frac{2}{3}} & 0 & 0 & 0 & -\frac{\sqrt{2}\omega^{2}}{3} & -\frac{\sqrt{5}}{3} & \frac{\omega}{3\sqrt{2}} & 0 & 0 & 0 
\end{smallmatrix}\right)}
\end{align}

\begin{align}
& \mathcal{M}_{2,z}= \nonumber \\
& {\footnotesize \left(\begin{smallmatrix} 
0 & 0 & 0 & 0 & \frac{1-\omega^{2}}{6} & 0 & \frac{1-\omega}{6} & 0 & 0 & 0 \\
0 & -\frac{1}{3} & -\frac{\sqrt{5}\omega}{3\sqrt{2}} & \frac{\omega^{2}}{6} & 0 & 0 & 0 & 0 & 0 & 0 \\
0 & -\frac{\sqrt{5}\omega^{2}}{3\sqrt{2}} & -\frac{1}{3} & -\frac{\sqrt{5}\omega}{3\sqrt{2}} & 0 & 0 & 0 & 0 & 0 & 0 \\
0 & \frac{\omega}{6} & -\frac{\sqrt{5}\omega^{2}}{3\sqrt{2}} & -\frac{1}{3} & 0 & 0 &  & 0 & 0 & 0 \\
\frac{1-\omega}{6} & 0 & 0 & 0 & 0 & -\frac{\omega^{2}}{\sqrt{10}} & -\frac{\omega}{2} & 0 & 0 & 0 \\
0 & 0 & 0 & 0 & -\frac{\omega}{\sqrt{10}} & 0 & -\frac{\omega^{2}}{\sqrt{10}} & 0 & 0 & 0 \\
\frac{1-\omega^{2}}{6} & 0 & 0 & 0 & -\frac{\omega^{2}}{2} & -\frac{\omega}{\sqrt{10}} & 0 & 0 & 0 & 0 \\
0 & 0 & 0 & 0 & 0 & 0 & 0 & \frac{1}{3} & \frac{\sqrt{5}\omega}{3\sqrt{2}} & -\frac{\omega^{2}}{6} \\
0 & 0 & 0 & 0 & 0 & 0 & 0 & \frac{\sqrt{5}\omega^{2}}{3\sqrt{2}} & \frac{1}{3} & \frac{\sqrt{5}\omega}{3\sqrt{2}} \\
0 & 0 & 0 & 0 & 0 & 0 & 0 & -\frac{\omega}{6} & \frac{\sqrt{5}\omega^{2}}{3\sqrt{2}} & \frac{1}{3}
\end{smallmatrix} \right)} \\
%
& \mathcal{M}_{2,+} = \nonumber \\ 
& {\footnotesize \left(\begin{smallmatrix} 
0 & \frac{\sqrt{2}\left(\omega-1\right)}{3} & 0 & \frac{\sqrt{2}\left(\omega^{2}-1\right)}{3} & 0 & 0 & 0 & 0 & 0 & 0 \\
0 & 0 & 0 & 0 & 0 & 0 & 0 & 0 & 0 & 0 \\
0 & 0 & 0 & 0 & 0 & 0 & 0 & 0 & 0 & 0 \\
0 & 0 & 0 & 0 & 0 & 0 & 0 & 0 & 0 & 0 \\
0 & \frac{\omega}{3\sqrt{2}} & \frac{\omega^{2}}{3\sqrt{5}} & \frac{1}{3\sqrt{2}} & 0 & 0 & 0 & 0 & 0 & 0 \\
0 & -\frac{\sqrt{5}\omega^{2}}{3} & -\frac{\sqrt{2}}{3} & -\frac{\sqrt{5}\omega}{3} & 0 & 0 & 0 & 0 & 0 & 0 \\
0 & \frac{1}{3\sqrt{2}} & \frac{\omega}{3\sqrt{5}} & \frac{\omega^{2}}{3\sqrt{2}} & 0 & 0 & 0 & 0 & 0 & 0 \\
\frac{\sqrt{2}\left(1-\omega^{2}\right)}{3} & 0 & 0 & 0 & \frac{\omega^{2}}{3\sqrt{2}} & -\frac{\sqrt{5}\omega}{3} & -\frac{\sqrt{2}}{3} & 0 & 0 & 0 \\
0 & 0 & 0 & 0 & \frac{4\omega}{3\sqrt{5}} & -\frac{\sqrt{2}}{3} & \frac{4\omega^{2}}{3\sqrt{5}} & 0 & 0 & 0 \\
\frac{\sqrt{2}\left(1-\omega\right)}{3} & 0 & 0 & 0 & -\frac{\sqrt{2}}{3} & -\frac{\sqrt{5}\omega^{2}}{3} & \frac{\omega}{3\sqrt{2}} & 0 & 0 & 0
\end{smallmatrix} \right)} 
\end{align}

\begin{align}
& \mathcal{M}_{3,z}= \nonumber  \\
& {\footnotesize \left(\begin{smallmatrix}
0 & 0 & 0 & 0 & -\frac{i}{2\sqrt{3}} & 0 & \frac{i}{2\sqrt{3}} & 0 & 0 & 0 \\
0 & -\frac{1}{3} & -\frac{\sqrt{5}\omega^{2}}{3\sqrt{2}} & 0 & 0 & 0 & 0 & 0 & 0 & 0 \\
0 & -\frac{\sqrt{5}\omega}{3\sqrt{2}} & -\frac{1}{3} & -\frac{\sqrt{5}\omega^{2}}{3\sqrt{2}} & 0 & 0 & 0 & 0 & 0 & 0 \\
0 & 0 & -\frac{\sqrt{5}\omega}{3\sqrt{2}} & -\frac{1}{3} & 0 & 0 & 0 & 0 & 0 & 0 \\
\frac{i}{2\sqrt{3}} & 0 & 0 & 0 & 0 & -\frac{1}{\sqrt{10}} & -\frac{1}{2} & 0 & 0 & 0 \\
0 & 0 & 0 & 0 & -\frac{1}{\sqrt{10}} & 0 & -\frac{1}{\sqrt{10}} & 0 & 0 & 0 \\
-\frac{i}{2\sqrt{3}} & 0 & 0 & 0 & -\frac{1}{2} & -\frac{1}{\sqrt{10}} & 0 & 0 & 0 & 0 \\
0 & 0 & 0 & 0 & 0 & 0 & 0 & \frac{1}{3} & \frac{\sqrt{5}\omega^{2}}{3\sqrt{2}} & -\frac{\omega}{6} \\
0 & 0 & 0 & 0 & 0 & 0 & 0 & \frac{\sqrt{5}\omega}{3\sqrt{2}} & \frac{1}{3} & \frac{\sqrt{5}\omega^{2}}{3\sqrt{2}} \\
0 & 0 & 0 & 0 & 0 & 0 & 0 & -\frac{\omega^{2}}{6} & \frac{\sqrt{5}\omega}{3\sqrt{2}} & \frac{1}{3}                                  
\end{smallmatrix} \right)} 
\end{align}

\begin{align}
& \mathcal{M}_{3,+}= \nonumber \\
& {\footnotesize \left(\begin{smallmatrix}
0 & \frac{\sqrt{2}\left(1-\omega^{2}\right)}{3} & 0 & \frac{\sqrt{2}\left(1-\omega\right)}{3} & 0 & 0 & 0 & 0 & 0 & 0 \\
0 & 0 & 0 & 0 & 0 & 0 & 0 & 0 & 0 & 0 \\
0 & 0 & 0 & 0 & 0 & 0 & 0 & 0 & 0 & 0 \\
0 & 0 & 0 & 0 & 0 & 0 & 0 & 0 & 0 & 0 \\
0 & \frac{\omega}{3\sqrt{2}} & \frac{1}{3\sqrt{5}} & \frac{\omega^{2}}{3\sqrt{2}} & 0 & 0 & 0 & 0 & 0 & 0 \\
0 & -\frac{\sqrt{5}\omega}{3} & -\frac{\sqrt{2}}{3} & -\frac{\sqrt{5}\omega^{2}}{3} & 0 & 0 & 0 & 0 & 0 & 0 \\
0 & \frac{\omega}{3\sqrt{2}} & \frac{1}{3\sqrt{5}} & \frac{\omega^{2}}{3\sqrt{2}} & 0 & 0 & 0 & 0 & 0 & 0 \\
\frac{\sqrt{2}\left(\omega-1\right)}{3} & 0 & 0 & 0 & \frac{\omega^{2}}{3\sqrt{2}} & -\frac{\sqrt{5}\omega^{2}}{3} & -\frac{\sqrt{2}\omega^{2}}{3} & 0 & 0 & 0 \\
0 & 0 & 0 & 0 & \frac{4}{3\sqrt{5}} & -\frac{\sqrt{2}}{3} & \frac{4}{3\sqrt{5}} & 0 & 0 & 0 \\
\frac{\sqrt{2}\left(\omega^{2}-1\right)}{3} & 0 & 0 & 0 & -\frac{\sqrt{2}\omega}{3} & -\frac{\sqrt{5}\omega}{3} & \frac{\omega}{3\sqrt{2}} & 0 & 0 & 0
\end{smallmatrix} \right)} \label{eq:Mplus3}
\end{align}

\bibliography{Kagome_S1.bib}
\end{document}